\documentclass[pss]{wiley2sp}
\usepackage{amsmath}
\usepackage{bm}
\usepackage{w-greek}

\begin{document}

\title{Scanning probe microscopy and spectroscopy of graphene on metals}

\titlerunning{SPM of graphene on metals}

\author{Yuriy Dedkov\textsuperscript{\textsf{\bfseries 1},\Ast}, Elena Voloshina\textsuperscript{\textsf{\bfseries 2}}, and Mikhail Fonin\textsuperscript{\textsf{\bfseries 3}}}

\authorrunning{Yu. Dedkov et al.}

\mail{e-mail \textsf{Yuriy.Dedkov@specs.com}, Phone: +49-30-467824-9339, Fax: +49-30-4642-083}

\institute{%
  \textsuperscript{1}\,SPECS Surface Nano Analysis GmbH, Voltastra\ss e 5, 13355 Berlin, Germany\\
  \textsuperscript{2}\,Institut f\"ur Chemie, Humboldt-Universit\"at zu Berlin, 10099 Berlin, Germany\\
  \textsuperscript{3}\,Fachbereich Physik, Universit\"at Konstanz, 78457 Konstanz, Germany}

\received{XXXX, revised XXXX, accepted XXXX} 
\published{XXXX} 


\abstract{
%
%
%
\abstcol{Graphene -- a two-dimensional (2D) material with unique electronic properties appears to be an ideal object for the application of surface science methods. Among them, a family of scanning probe microscopy methods (STM, AFM, KPFM) and the corresponding spectroscopy add-ons provide information about structure and electronic properties of graphene on the local scale (from $\mu$m to atoms). This review focuses on the recent applications of these microscopic/spectroscopic methods for the investigation of graphene on metals (interfaces, intercalation-like systems, graphene nanoribbons and quantum dots, etc). It is shown that very important information about interaction strength at the graphene-metal interfaces as well as about modification of the electronic spectrum of graphene at the Fermi level can be obtained on the local scale. The combination of these results with those obtained by other methods and comparison with recent theoretical data demonstrate the power of this approach for the investigation of the graphene-based systems.}}
%
%
\titlefigure[height=5.5cm]{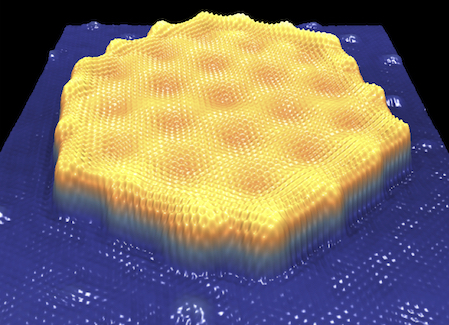}
\titlefigurecaption{STM image of a graphene island (quantum dot) on Ir(111).}

\maketitle

\section{Introduction}

The discovery of the fascinating electronic transport properties of graphene~\cite{Novoselov:2004a,Novoselov:2005,Zhang:2005}, a two-dimensional (2D) allotrope form of carbon, has attracted intense attention of materials science and solid state physicists~\cite{CastroNeto:2009,Geim:2009}, because of fundamental physics as well as the device application perspectives in electronics and spintronics. First exfoliated and identified in 2004~\cite{Novoselov:2004a,Novoselov:2005,Zhang:2005}, graphene was actually known for many years in the surface science community as a graphitic ``dead'' layer, which poisons the catalitic activity of metal surfaces~\cite{Hagstrom:1965vh,May:1969uj,Grant:1970,Land:1992}. However, the interest in the graphene/metal systems experienced a strong revival since it has been realized that the synthesis of graphene on metal surfaces is the most prospective method to obtain high quality large-area graphene samples for further applications~\cite{Li:2009,Bae:2010,Tao:2012tc}.

Beyond the possible technological relevance, the investigation of continuous graphene layers as well as graphene nanostructures on metal surfaces turned out to be challenging and exciting from both experimental and theoretical point of view. One of the most prominent examples in this regard addresses the nature of bonding at the graphene/metal interface, the problem, which was discussed in a large series of publications (see \textit{e.\,g.} Refs.~\cite{Wintterlin:2009,Voloshina:2012c,Dedkov:2012book}), having the question open for long time -- how relatively weak interaction (in the order of $50-150$\,meV/C-atom) can lead to strong modifications in the electronic structure of graphene? Only recently this problem was considered in a complex approach and the universal model, describing the interaction in the whole graphene/metal interface family, was proposed~\cite{Voloshina:2014jl}. The search for the routes to minimize this interaction especially aims at the preparation of the graphene nanoribbons~\cite{Cai:2010,Ruffieux:2012et,Linden:2012fu,Talirz:2013es,vanderLit:2013jf,Zhang:2014em} or epitaxial nano sized islands with different edge terminations~\cite{Subramaniam:2012fp,Altenburg:2012kp,Li:2013ji,Leicht:2014jy}, where the low-dimensional effects such as a band-gap width quantization and the edge-induced magnetism are expected. Due to the truly 2D character of the crystallographic and the electronic structure of graphene and the localisation of the interesting phenomena at the small scale, the scanning probe microscopy (SPM) methods, i.e. scanning tunnelling microscopy (STM) and atomic force microscopy (AFM) in combination with the corresponding spectroscopy techniques (STS and AFS), can provide important information about the properties of graphene nanostructures at the nanometer- and atomic-scale.

This feature article gives a short review of the recent STM and AFM experiments on continuous graphene layers and  graphene nanostructures prepared on metallic surfaces. The presented results are linked to the available theoretical data providing an important input for the understanding of the observed phenomena.

\begin{figure}[t]
\includegraphics*[width=\linewidth]{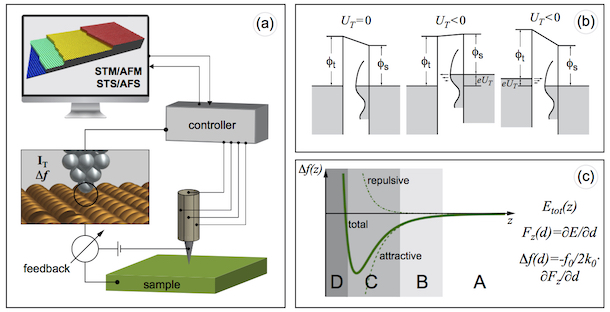}
\caption{(a) Principal scheme of an SPM experiment. (b) Energy scheme of the STM experiment for different bias voltages applied between the tip and the sample. (c) Frequency shift of the sensor ($\Delta f$) as a function of the distance between tip and sample ($z$) expressed from the total interaction energy ($E_{tot}$); A: no interaction, B: long-range vdW and electrostatic forces, C: short-range chemical forces giving atomic contrast in AFM, D: short-range repulsive interaction.}
\label{SPM_methods}
\end{figure}

\section{Scanning probe microscopy and spectroscopy}

\subsection{STM and STS}

In STM experiments the sharp conductive tip is moved above the sample surface in $x$ and $y$ directions [Fig.~\ref{SPM_methods}(a)]. In constant current STM measurements the tunnelling current $I_T$, which exponentially depends on the distance $z$ between tip and sample [$I_T\propto exp(-kz)$], is kept constant by a feedback loop adjusting the actual $z$ position, which is recorded to obtain a three-dimensional topography map. The obtained constant current image represents the integral local density of states (LDOS) of the sample surface, $I_T(x,y,z,V)\propto \int \rho(x,y,z,E)dE$, where the integration is performed in the energy range between the Fermi levels ($E_F$) of tip and sample, when they differ by the value $eU_T$, where $U_T$ is the bias voltage [Fig.~\ref{SPM_methods}(b)]. Differentiation of the $I(x,y,z,U_T)$ with respect to the bias voltage (can be performed with lock-in technique) gives a direct information about the local density of states. Such measurements carried out in the scanning mode at different bias voltages allow to observe the so-called electron density standing waves of different periodicities that can be used to obtained the electron dispersion relation, $E(k)$, of the surface electronic states.

\subsection{AFM and AFS}

In the modern AFM experiments, frequency-modulated AFM (FM-AFM), the oscillating at high resonance frequency scanning sensors (based on quartz tuning forks or length-extensional resonators) are widely used~\cite{Giessibl:2003aaa,Giessibl:2011fw}. In this case the sensitivity of the method can be increased dramatically. Here, a tiny conductive tip connected to the resonator approaches the sample surface and the interaction between them leads to the change of the resonance frequency by value $\Delta f$: $F_z(d)=\partial E/\partial d$, $\Delta f(d)=-f_0/2k_0\cdot\partial F_z(d)/\partial d$, where $E(d)$ and $F_z(d)$ are the interaction energy and the vertical force between a tip and the sample, respectively, $f_0$ and $k_0$ are the resonance frequency and the spring constant of the sensor [Fig.~\ref{SPM_methods}(c)]. The detected frequency shift, $\Delta f$, is used in the feedback loop and the corresponding ``topography'' map can be obtained in such AFM measurements. The resulting interaction between the sample and the tip is the sum of the long-range electrostatic ($F_{el}$) and van der Waals ($F_{vdW}$) contributions and the short-range chemical ($F_{chem}$) interaction. Similar to the STM experiments, $F_{chem}$ at short distances between tip and sample produces the atomic contrast in AFM experiments and as was shown in a series of the recent works the observed imaging contrast depends on several factors, \textit{e.\,g.}, distance and the setting point for the measurements [Fig.~\ref{SPM_methods}(c)]. The locally measured $F_z(d)$ curves can be used for the calculation of the tip-sample interaction forces via methods developed by Giessibl~\cite{Giessibl:2001ig} or Sader and Jarvis~\cite{Sader:2004kt}.

As an \textit{add-on} to AFM, the method of the Kelvin-probe force microscopy (KPFM) allowing to obtain the local distribution of the electrostatic potential was developed~\cite{Nonnenmacher:1991,Melitz:2011}. In this method the topographic measurements are performed in AFM mode and at the same time the DC ($U_{DC}$) and AC voltages are applied between tip and sample. Here the lock-in technique is used and the first harmonic in the output signal, which is proportional to the difference between $U_{DC}$ and the local contact potential difference ($U_{LCPD}$) between conductive tip and sample, is nullified allowing to extract the $U_{LCPD}(x,y)$ map of the sample surface. 

\begin{figure}[t]
\includegraphics*[width=\linewidth]{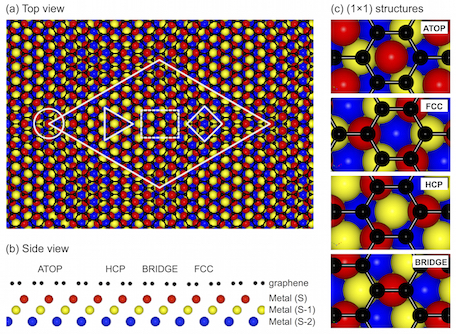}
\caption{(a,b) Top and side views of the crystallographic model of the graphene moir\'e structure on the close-packed (111) metallic surface, ($10\times10$)graphene/($9\times9$) Metal(111). (c) Structures of the local high-symmetry positions of the graphene/metal interfaces.}
\label{structures}
\end{figure}

\subsection{3D AFM/STM}

The 3D STM/AFM measurements can be performed in two ways. In the first case the distance between tip and sample is fixed with the feedback loop completely switched off and the respective $I(x,y)$ and $\Delta f(x,y)$ maps are simultaneously collected. If the distance between tip and sample is varied with the regular steps, then the dense 3D data sets can be collected, $I(x,y,z)$ and $\Delta f(x,y,z)$. In the second approach the single $I(z)$ and $\Delta f(z)$ curves are measured on the $(x,y)$-grid and then they are combined in the 3D sets. Both measurement schemes are time consuming and here the special attention should be paid to the possible creep and thermal drifts of the piezo-drive and the AFM sensor as well as to the post-correction of the obtained data. Different schemes of 3D STM/AFM measurements and their (dis)advantages are discussed in Refs.~\cite{Abe:2007jm,Sugimoto:2008bs,Albers:2009ig,Sugimoto:2012dc}. 

\section{Theoretical approaches: DFT, models, \textit{etc.}}

Considering the arrangement of the graphene layer on a metal surface, whether lattice-matched or lattice-mismatched system, one can identify several high-symmetry stacking positions for carbon atoms in the layer. (Note: Different notation are used in the literature to mark these positions). They are:
\vspace{-0.15cm}
\begin{itemize}
\item[(a) ] ATOP (\textit{hcp-fcc}) position, where carbon atoms surround the metal atom of the top layer and are placed in the \textit{hcp} and \textit{fcc} hollow sites of the Metal(111) stack above (S-1) and (S-2) Metal-layers, respectively [Circle in Fig.~\ref{structures} (a) and the respective arrangements in Fig.~\ref{structures} (b) and (c)];
\item[(b) ] FCC (\textit{top-hcp}) position, where carbon atoms surround the \textit{fcc} hollow site of the Metal(111) surface and are placed in the \textit{top} and \textit{hcp} hollow positions of the Metal(111) stack above (S) and (S-1) Metal-layers, respectively [Rhombus in Fig.~\ref{structures} (a) and the respective arrangements in Fig.~\ref{structures} (b) and (c)];
\item[(c) ] HCP (\textit{top-fcc}) position, where carbon atoms surround the \textit{hcp} hollow site of the Metal(111) surface and are placed in the \textit{top} and \textit{fcc} hollow positions of the Metal(111) stack above (S) and (S-2) Metal-layers, respectively  [Triangle in Fig.~\ref{structures} (a) and the respective arrangements in Fig.~\ref{structures} (b) and (c)];
\item[(d) ] BRIDGE-position, where carbon atoms are bridged by the Metal atom of the (S) layer [Rectangle in Fig.~\ref{structures} (a) and the respective arrangements in Fig.~\ref{structures} (b) and (c)].
\end{itemize}

The vast majority of computational studies of graphene on metals are currently performed using first-principles electronic structure methods based on density functional theory (DFT). While density functional theory itself is capable of providing the exact solution to the Schr\"odinger equation, including long-range correlations - the dispersion, the approximations made in the DFT functionals of all types yield a rather unsatisfactory description of the intermolecular interactions. Therefore when studying graphene/metal interfaces, van-der-Waals bound systems, one usually employs \textit{a posteriori} dispersion correction schemes, such as DFT-D method of Grimme~\cite{Grimme:2004,Grimme:2006,Grimme:2010} or Tkachenko-Scheffler method~\cite{Tkatchenko:2009}, and the obtained results are usually in reasonable agreement with experiment. Unlike these post-DFT corrections, an \textit{a priori} way consists in utilisation of non-local correlation functionals that approximately accounts for dispersion interactions, the so-called vdW-DF\cite{Dion:2004,Klimes:2009ei}. Although the latter approach is expected to give more accurate results, it is shown to predict irrationally weak binding of graphene on metals surfaces for all studied cases~\cite{Vanin:2010cr}. 

\begin{figure}[t]
\includegraphics*[width=\linewidth]{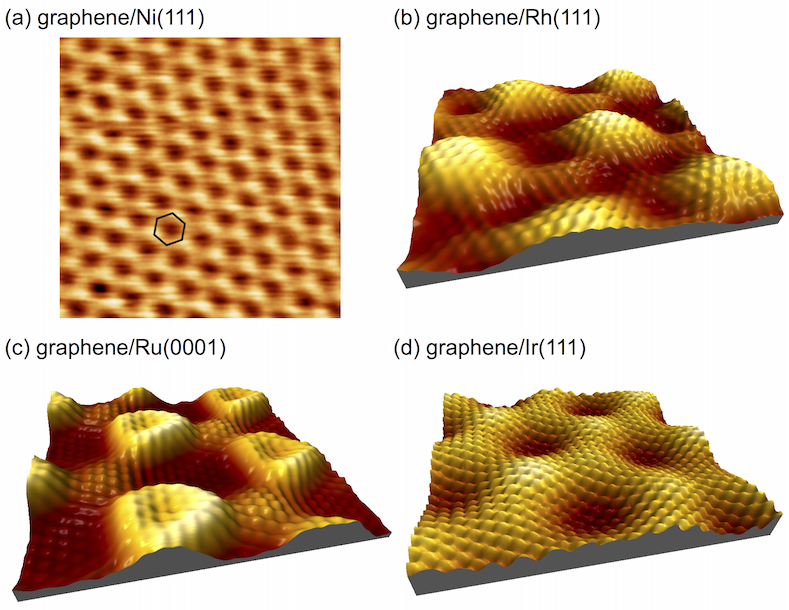}
\caption{STM images of (a) lattice matched graphene/Ni(111) ($2.5\times2.5\mathrm{nm}^2$, $U_T=2\mathrm{mV}$, $I_T=48\mathrm{nA}$) and lattice mismatched systems: (b) graphene/Rh(111) ($5.4\times5.4\mathrm{nm}^2$, $U_T=300\mathrm{mV}$, $I_T=1.6\mathrm{nA}$), (c) graphene/Ru(0001) ($5.7\times5.7\mathrm{nm}^2$, $U_T=300\mathrm{mV}$, $I_T=1.6\mathrm{nA}$), and (d) graphene/Ir(111) ($4.5\times4.5\mathrm{nm}^2$, $U_T=300\mathrm{mV}$, $I_T=1.6\mathrm{nA}$).}
\label{STM_grNiRhRuIr}
\end{figure}

A very important analysis of the surface atomic and electronic structure is the simulation of STM images. Among early theoretical models to describe atomically resolved STM images Tersoff and Hamann approach~\cite{Tersoff:1985} has found particular attention since it relates the observed contrast to a conceptually simple quantity of the surface, namely the local density of states of valence or conduction electrons. While it provides a simple interpretation of the STM results, this model  ignores the influence of the tunnelling barrier, the structural and electronic properties of the tip and its interaction with the sample surface. However, more exact theoretical formulations of the STM problem (for a review, see Ref.~\cite{Hofer:2003wk}) were found to be much more time-consuming. At the same time, in all studied graphene/metal cases the Tersoff-Hamann approach was found to be rather reliable.

While STM can achieve atomic resolution, it probes the surface electronic structure of the sample in a mode that does not provide a direct interpretation of the atomic structure. First-principles simulations of NC-AFM can enhance the interpretation of experimental measurements; however, such simulations remain a challenge because they involve calculations of the sample together with an atomic model of the AFM tip (see e.g. Ref.~\cite{Kantorovich:2006ca}). Hence, the simulation can be very laborious. An efficient scheme to simulate NC-AFM images using inputs from first-principles calculations of the sample only without explicit modelling of the AFM tip was proposed by Chan \textit{et al.}~\cite{Chan:2009jb}. Both of these approaches were found to work reasonably well for the considered graphene/metal systems. 

\section{Graphene on metals: SPM/DFT of lattice match-ed and lattice-mismatched systems}

The history of the surface science studies of graphene on metals can be traced back to the middle of the 60s. However, the first STM experiments on graphitic layers on metal surfaces were performed in the beginning of the 90s when graphitic layers on Ni(111)~\cite{Klink:1995} and Pt(111)~\cite{Land:1992} were studied. It is worth to mention that already at that time the graphene moir\'e structures of different periodicities were identified on Pt(111).

The simplest, from the structural point of view, graphene-metal interface is graphene/Ni(111) or graphene/ Co(0001) as the difference between lattice constants of these surfaces and graphene is below $2$\%. Both systems were intensively studied by means of STM~\cite{Dedkov:2008a,Eom:2009,Varykhalov:2009,Dedkov:2010jh,Lahiri:2011iu,Dzemiantsova:2011bv,Jacobson:2012gv,Jacobson:2012be,Bianchini:2014faa,Prezzi:2014bk}. These studies demonstrate that a graphene layer deposited by means of chemical vapour deposition (CVD) from hydrocarbons at optimal experimental conditions forms a commensurate $(1\times1)$ structure on top of Ni and Co close-packed surfaces [Fig.~\ref{STM_grNiRhRuIr}(a)]. In this case, according to DFT calculations (with and without inclusion of the dispersive vdW interaction)~\cite{Bertoni:2004,Weser:2011,Voloshina:2011NJP,Dzemiantsova:2011bv}, graphene is adsorbed in the HCP (\textit{top-fcc}) configuration on Ni(111) or Co(0001). However, as shown within the same theoretical calculations, the energy difference between all high-symmetry stackings is rather small (below $100$\,meV) that can lead to the appearance of the moir\'e-like structures even for these lattice-matched interfaces~\cite{Dedkov:2010jh,Bianchini:2014faa}.

The interesting class of graphene-metal systems is obtained when graphene is prepared on the close-packed surfaces of $4d$ or $5d$ metals. In this case, the existing lattice mismatch leads to the formation of the so-called moir\'e graphene-metal structures of different periodicities. These systems were in the focus of the intensive STM investigations and the representative examples are shown in Fig.~\ref{STM_grNiRhRuIr}(b-d). It was found that the apparent corrugation of graphene-moir\'e strongly depends on the imaging bias voltage~\cite{NDiaye:2008qq,VazquezDeParga:2008,Stradi:2011be,Stradi:2012hw,Voloshina:2012a,Voloshina:2013dq} and at some tunnelling conditions inversion of the imaging contrast is observed. 

\begin{figure}[t]
\includegraphics*[width=\linewidth]{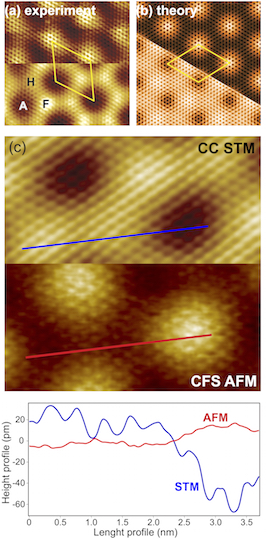}
\caption{(a,b) Experimental ($6\times6\mathrm{nm}^2$) and theoretical STM images for graphene/Ir(111), respectively. In the experimental data the bias voltage was changed from $-1.8$\,V (top) to $-0.3$\,V (bottom) during scanning. In the simulated images the respective integration energy ranges between $E_F$ and $eU_T$ were used. (c) Combined STM/AFM imaging of graphene/Ir(111) when scanning mode was changed ``on-the-fly'' during scanning ($5\times5\mathrm{nm}^2$, $U_T=30\mathrm{mV}$, $I_T=1\mathrm{nA}$, $\Delta f=-475\mathrm{mHz}$). The lower panel shows the corresponding height profiles for STM and AFM.}
\label{STM_grIr_bias}
\end{figure}

The representative example is shown in Fig.~\ref{STM_grIr_bias}(a) where bias voltage was changed during STM imaging of graphene/Ir(111) from $U_T=-0.3$\,V (bottom) to $U_T=-1.8$\,V (top)~\cite{Voloshina:2013dq,Dedkov:2014di}. Here at the typical tunnelling voltages used in STM, the graphene/Ir(111) system is imaged in the so-called \textit{inverted} contrast when topographically highest places of the moir\'e structure (ATOP=A) are imaged as dark areas and the lowest places are images as bight (FCC=F and HCP=H) [Fig.~\ref{STM_grIr_bias}(a), bottom]. Beyond the change of the imaging contrast, the apparent corrugation is also changed from $\approx0.32$\,\AA\ ($U_T=-0.3$\,V) to $\approx0.14$\,\AA\ ($U_T=-1.8$\,V). This effect was successfully explained theoretically where graphene/Ir(111) was treated on the DFT-D2 level. These calculations correctly describe the crystallographic structure of graphene/Ir(111) with the corrugation of graphene of $31$\,pm (PBE-D2)~\cite{Voloshina:2013dq} or $35$\,pm (vdW-DFT)\cite{Busse:2011} and reproduce the dependence of the imaging contrast as well as the corrugation in STM experiment on the bias voltage~\cite{Voloshina:2013dq}. The effect of the contrast inversion was assigned to the so-called interface states of graphene/Ir(111) formed as a result of the overlap of graphene $p_z$ and Ir $5d_{z^2}$ orbitals. The value of the graphene corrugation of $43\pm9$\,pm was obtained in the LEED-IV experiments supported by the AFM results (with CO-terminated tip), which yield a value of corrugation of $42-56$\,pm~\cite{Hamalainen:2013jj}. Contrary to these results, the much larger corrugation of $0.6$\,\AA\ and $1.0$\,\AA\ for $0.39$\,ML and $0.63$\,ML graphene on Ir(111), respectively, was obtained from an x-ray standing wave (XSW) experiments~\cite{Busse:2011} and this large corrugation was assigned to the increasing of the number of lines of wrinkles on the surface. The similar approach was used for the explanation of the effect observed during imaging of graphene/Ru(0001)~\cite{VazquezDeParga:2008,Stradi:2011be,Stradi:2012hw}.

\begin{figure}[t]
\includegraphics*[width=\linewidth]{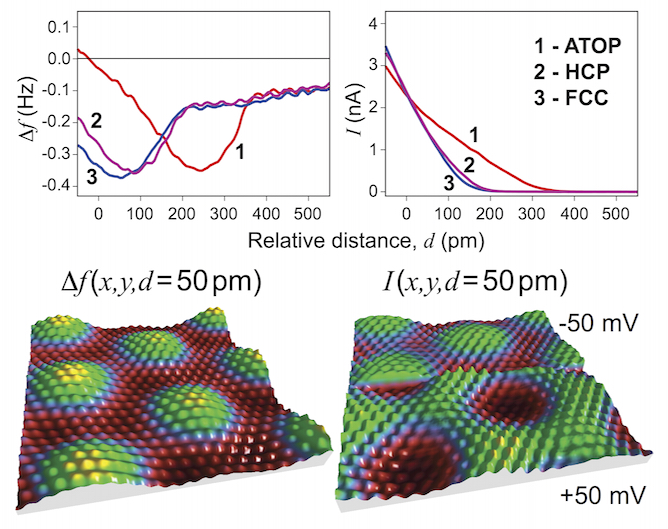}
\caption{Upper panel: $\Delta f(z)$ and $I(z)$ curves measured for different high-symmetry positions of graphene/Ir(111). Lower panel: simultaneously collected pure constant height $\Delta f(x,y)$ and $I(x,y)$ maps ($3.9\times3.9\mathrm{nm}^2$) at the relative distance between tip and sample of $d=50$\,pm. The bias voltage was changed from $+50$\,mV to $-50$\,mV in the middle of the scanning area.}
\label{grIr_CHAFM}
\end{figure}

One of the most interesting examples of graphene growth on a $3d$ metal surface, which also delivers an unusual moir\'e structure symmetry, is the graphene/Fe(110) system~\cite{Vinogradov:2012fg}. The preparation of graphene on top of a Fe(110) film, although being challenging due to the possible formation of different carbidic phases, has been successfully realized by a standard preparation method relying on thermal dissociation of the C$_3$H$_6$ precursor at the hot metal surface. Corresponding STM images of the graphene surface show a pronounced corrugation yielding a periodic wavy pattern, with the lateral distance of 4\,nm between the waves running along the [001] direction of the underlying substrate. This very special moir\'e structure stems from a unique combination of the lattice mismatch at the interface and the strong graphene/substrate interaction as shown by the accompanying DFT calculations~\cite{Vinogradov:2012fg}.

Application of the fast and reliable scanning probe sensors (Kolibrisensor\texttrademark from SPECS or QPlus from Omicron) allows to perform \textit{simultaneous} STM/AFM imaging of the same sample area and delivers interesting results for the graphene-metal moir\'e structures. In AFM the interaction between tip and sample is the sum of the long-range vdW and electrostatic forces and the short-range chemical forces. The latter allows to perform the atomically resolved imaging in AFM. In this case the imaging contrast in AFM depends on the distance between tip and sample and the corresponding $\Delta f$ value used in the feedback loop of SPM. In the recent experiments performed on graphene/Ir(111) the imaging mode was changed between STM and AFM (in the attractive regime) during scanning that allows to carefully trace the topographic contrast in this system [Fig.~\ref{STM_grIr_bias}(c)]. One can clearly see that the imaging contrast between CC STM and CFS AFM is inverted: STM demonstrates \textit{inverted} contrast and AFM performed in the attractive regime far from the minimum point for the $\Delta f$ signal gives true \textit{direct} contrast. The corresponding corrugation of the moir\'e structure is $100$\,pm and $24$\,pm in STM and AFM, respectively. At the same time the constant-height measurements (Fig.~\ref{grIr_CHAFM}) performed for the repulsive part of the interaction curve of the tip and sample ($d=50$\,pm) show the stronger repulsive interaction between ATOP place of graphene/Ir(111) and the tip compared to other regions of this structure. Change of the sign of the bias voltage during such measurements leads to the inversion of the sign of the tunnelling current (value of the current depends on the local density of states in the particular energy range around $E_F$), but it does not influence the $\Delta f$ channel as the electrostatic force between tip and sample depends as $F_{el}\propto -U_T^2$ and no ``cross-talk'' between $\Delta f$ and $I$ channels is detected.

\begin{figure}[t]
\includegraphics*[width=0.9\linewidth]{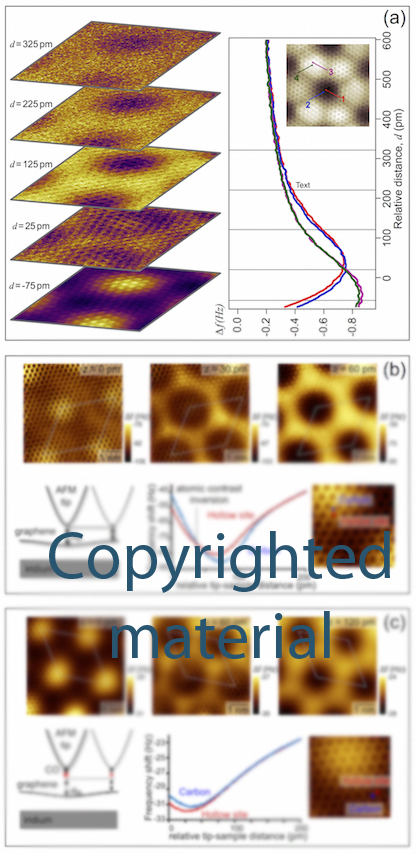}
\caption{(a) Constant height $\Delta f(x,y)$ maps ($3.5\times2.9\mathrm{nm}^2$) extracted from the 3D data set for graphene/Ir(111) at different relative distances between tip and sample (left-hand side). The atomically resolved $\Delta f(z)$ curves are shown in the right-hand panel for different places marked in the STM image (inset). (b) and (c) demonstrate the difference in the AFM and AFS imaging of graphene/Ir(111) with metallic and CO-terminated tip, respectively. Data in (b) and (c) are taken from Ref.~\cite{Boneschanscher:2012bg} with permission.}
\label{grIr_spectroscopy}
\end{figure}

The 3D force and current spectroscopy, $\Delta f(z)$ and $I(z)$, performed on the grid for graphene/Ir(111) demonstrates the inversion of the imaging contrast for the $\Delta f$ channel in the constant height AFM measurements~\cite{Voloshina:2013dq,Dedkov:2014di,Boneschanscher:2012bg}. The results of such measurements are presented in Fig.~\ref{grIr_spectroscopy}(a) where on the left-hand side the $\Delta f(z)$ curves acquired at the different places of the graphene moir\'e structure on Ir(111) are shown. One can clearly see that several so-called \textit{crossing points} for the $\Delta f$ curves appear on this plot that indicates that inversion of the imaging contrast in CH AFM will be observed on the moir\'e-cell scale as well as on the atomic scale. The corresponding cuts, $\Delta f(x,y,d)$, extracted from the 3D data sets are shown on the left-hand side of Fig.~\ref{grIr_spectroscopy}(a), where such effect of the contrast inversion as a function of the distance between oscillating tip and sample is clearly observed~\cite{Voloshina:2013dq,Dedkov:2014di}.

The interesting effect was observed in the AFM imaging and spectroscopy of graphene/Ir(111) when clean metallic and a CO-passivated tip were used in the experiment [Fig.~\ref{grIr_spectroscopy}(b,c)]~\cite{Boneschanscher:2012bg}. Imaging with the reactive metallic Ir tip shows that atomic contrast on graphene/Ir(111) can be obtained in the attractive regime, which is changed to the repulsive atomic imaging at the shorter distances. Here the inversion of the imaging contrast is observed at the large (moir\'e lattice) as well as at the atomic scale [Fig.~\ref{grIr_spectroscopy}(b)]. If a non-reactive tip is used (CO-terminated Ir tip) [Fig.~\ref{grIr_spectroscopy}(c)], then imaging of the structure is different: at large distances only ATOP places can be clearly resolved; at shorter distance the Pauli repulsion between graphene and CO leads to the appearance of the honeycomb structure formed from carbon atoms; inversion of the contrast is observed only at the large scale of the moir\'e unit cell and not at the atomic scale, which is fully consistent with the $\Delta f(z)$ curves presented in Fig.~\ref{grIr_spectroscopy}(c).

In the similar 3D AFM experiments on the graphene/ Ru(0001) system, the mechanical properties of the graphene nanodomes, which are of the height of $1.1$\,\AA\ were investigated~\cite{Koch:2013bk}. The CFS AFM images of this system demonstrate the vertical reversible deformation of the hill places of the moir\'e structure. The systematic measurements and comparison with theory allow to estimate the stiffness of the graphene nanomembrane to the value of $k_{dome}=43.6\pm0.5$\,N/M with the resonance frequency of $\sim2$\,THz and it was suggested that such nanodomes can be used as nanoelectromechanical resonators.

Graphene moir\'e structures on the lattice-mismatched metal surfaces demonstrate also the local variation of the chemical potential. For example, on the strongly corrugated graphene/Rh(111) and graphene/Ru(0001) surfaces the periodic modulation of the local work function is $0.22$\,eV~\cite{Wang:2010ky} and $0.52$\,eV~\cite{Borca:2010jj}, respectively, as obtained in the spectroscopic measurements of the image potential states with STM above different places of graphene moir\'e. These results were confirmed by DFT calculations for the \textit{strongly} interacting graphene/Ru(0001) system with large corrugation~\cite{Wang:2010jw} and explained by the different local interaction strength for topographically high ATOP and low FCC and HCP positions. For the \textit{weakly} interacting graphene on Ir(111), which has small corrugation of $\approx0.3$\,\AA\ with the mean distance between graphene and metal surface of $3.3$\,\AA, the variation of the local work function in graphene moir\'e is $\approx100$\,meV as measured from $I(z)$ spectroscopy data in STM~\cite{Altenburg:2014cq}. At the same time the KPFM measurements give a lower value of $35$\,meV~\cite{Dedkov:2014di} that can be connected with the underestimation of this value in KPFM measurements on the nm-scale. DFT calculations for this system yield value of $56$\,meV.

\section{Properties of graphene-metal-based systems: adsorption and intercalation}

Adsorption of graphene on the metal surfaces modifies the electronic spectrum of graphene-related valence band states around $E_F$. There are several ways allowing to tailor the properties of the graphene/metal interface with the aim to control electronic and magnetic properties of a graphene layer. 

\begin{figure}[t]
\includegraphics*[width=\linewidth]{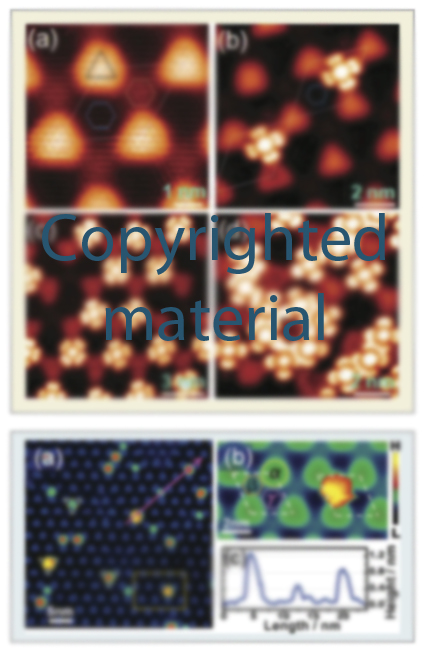}
\caption{Upper panel: STM images of FePc molecules adsorbed on graphene/Ru(0001) at varying coverage. Data are taken from Ref.~\cite{Zhang:2011ky} with permission. Lower panel: (a) STM topographic image of Pt NCs on graphene/ Ru(0001). (b) Zoom in the rectangular area in (a). (c) Line profile of three clusters along the purple arrow in (a), showing typical heights of $1.27$, $0.5$, and $0.88$\,nm. Data are taken from Ref.~\cite{Pan:2009} with permission.}
\label{STM_gr-met-ads}
\end{figure}

Adsorption of different species has the aim to tailor the doping level of graphene as well as, in the case of the lattice-mismatched graphene/metal interfaces, to create the ordered arrays of molecules~\cite{Zhang:2011ky,Zhang:2012il,Yang:2012kq,Li:2012in,Hamalainen:2012bx,Jarvinen:2014va} or metallic clusters~\cite{NDiaye:2009a,Pan:2009,Rusponi:2010,VoVan:2011ia,Papagno:2012hl,Knudsen:2012ei} where properties of the whole system can be modelled on the basis of the single element. As proposed, such systems can be used in the future catalysis elements or a parts of the future information storage devices. For example, it was shown that the existence of the lateral dipole moment in the graphene/Ru(0001) moir\'e structure leads to the site-selective adsorption of FePc and pentacene molecules on this structure and formation of the ordered molecular arrays [Fig.~\ref{STM_gr-met-ads} (upper panel)]~\cite{Zhang:2011ky}. This effect was attributed to the variation of the local work function along the graphene moir\'e as discussed earlier.

Adsorption of metallic clusters on top of the graphene/ metal moir\'e structure was studied in a series of experimental and theoretical works. It was shown that for the strongly corrugated graphene/Ru(0001) system adsorption of Pt favours to the FCC places of moir\'e [Fig.~\ref{STM_gr-met-ads} (lower panel)]~\cite{Pan:2009}, where as in the case of the weakly buckled graphene on Ir(111) adsorption of Pt and Ir clusters appears in HCP places~\cite{NDiaye:2009a,Ndiaye:2006}. In the latter case adsorption of Ir clusters leads to the local rehybridization of carbon orbitals from $sp^2$ to $sp^3$~\cite{Feibelman:2009} that was confirmed later in the photoemission experiments, where anisotropy of the Dirac cone was observed around the $K$ point of the graphene Brillouine zone and the increasing of the width of the replica-bands-induced mini-gaps compared to graphene/Ir(111)~\cite{Rusponi:2010}. The further modification of this system by doping with Na atoms leads to the drastic changes in the spectrum of graphene and opening of a large gap of $740$\,meV directly at the Dirac point of graphene~\cite{Papagno:2012hl}.

As has been discussed earlier the adsorption of a graphene layer on metal leads to the strong modification of the spectrum of the graphene-derived states in the vicinity of $E_F$: linear dispersion is altered via renormalisation of the effective mass of the carriers as well as the symmetry violation for the carbon sublattices in graphene might open the energy gap at the Dirac point. The most promising way to breake this ``strong'' interaction between graphene and metal is intercalation of different species, like the atoms of metals, gases or big molecules. This method was initially developed for graphite crystals where different substances were placed (intercalated) between carbon layers that changes the properties of graphite substantially~\cite{Dresselhaus:2002}.

\begin{figure}[t]
\includegraphics*[width=\linewidth]{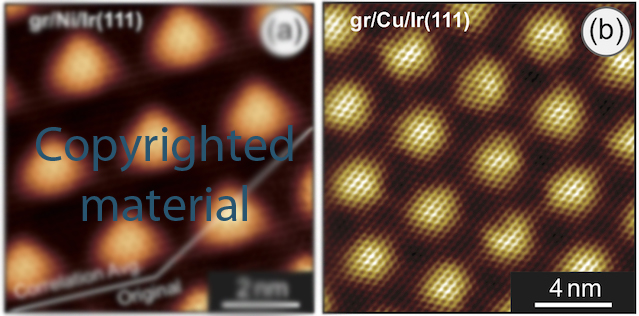}
\caption{STM images of (a) the graphene/Ni/Ir(111) ($U_T=50\mathrm{mV}$, $I_T=35\mathrm{nA}$)~\cite{Pacile:2013jc} and graphene/Cu/Ir(111) ($U_T=300\mathrm{mV}$, $I_T=1.6\mathrm{nA}$)~\cite{Vita:2014aa} intercalation-like systems. Data are reproduced with permission.}
\label{STM_grNiCuIr111}
\end{figure}

Recently the effect of intercalation of Co, Ni and Cu in graphene/Ir(111) was studied~\cite{Pacile:2013jc,Vita:2014aa,Decker:2013ch} [Fig.~\ref{STM_grNiCuIr111}(a,b)].  As was found, in all cases the pseudomorphic layers of intercalated metal (Co, Ni or Cu) were formed at the interface between graphene and metal. 

In case of the graphene/Ni(Co)/Ir(111) system~\cite{Pacile:2013jc,Decker:2013ch} [Fig.~\ref{STM_grNiCuIr111}(a)] the crystallographic structure of graphene was found to be similar to the one for the strongly buckled graphene layer on Ru(0001); the closest distance between graphene and Ni (Co) layer is $1.94$\,\AA\ ($2.02$\,\AA) and corrugation is $1.51$\,\AA\ ($1.27$\,\AA) compared to the same values of $2.195$\,\AA\ and $1.195$\,\AA\ for graphene/Ru(0001)~\cite{Stradi:2011be}. It was found that intercalation of Ni leads to the complete redistribution of the valence band states of graphene and graphene behaves as ``strongly'' interacting with metal and its properties (doping level as well as the formation of interface hybrid states) are determined by the FCC and HCP ``strongly'' interacting places of the graphene/Ni/Ir(111) system. In the consequent spin-polarised STM experiments with ferromagnetic scanning tip and in the presence of the magnetic field it was found that the Co islands have out-of-plane magnetic anisotropy (similar to Co/Ir(111)~\cite{Bickel:2011fg}); however the coercive field of the graphene/Co/Ir(111) was found to be higher than $\sim4.5$\,T, which is much larger compared to the one of $\approx3$\,T for Co/Ir(111)~\cite{Bickel:2011fg}. Comparison of the experimental results with DFT calculations shows that FCC and HCP carbon places which are closer to the Co layer are coupled antiferromagnetically to Co with magnetic moment of $-1.36\mu_B$/graphene-unit-cell, whereas for the ATOP places, where interaction between graphene and Co is weaker, the coupling between Co and graphene is ferromagnetic with magnetic moment of $+0.14\mu_B$/graphene-unit-cell.

The situation for graphene/Cu/Ir(111) is similar from the microscopy point of view~\cite{Vita:2014aa}[Fig.~\ref{STM_grNiCuIr111}(b)]. The pseudomorphic growth of Cu was found in STM experiments. At the same time a graphene layer on Cu/Ir(111) becomes less corrugated with the height difference of $0.229$\,\AA\ comapred to $0.307$\,\AA\ for graphene/Ir(111). Also the distance between graphene layer and metallic substrate is reduced after Cu intercalation from $3.581$\,\AA\ to $3.122$\,\AA\ for the ATOP positions. It is interesting to note that despite the relatively ``weak'' interaction between graphene and Cu/Ir(111) the electronic structure of graphene is significantly modified. Firstly, with respect to STM imaging, the inversion of the imaging contrast, characteristic for graphene/Ir(111), was not detected and graphene/Cu/Ir(111) is always imaged in the \textit{true} contrast: topographically highest places (ATOP) are imaged as bright spots that can be explained by similar carbon-projected local DOS around $E_F$. The electronic spectrum of graphene has an energy gap directly at $E_D$ (similar to graphene on Cu(111), Cu/Ni(111), Ag/Ni(111), Au/Ni(111)~\cite{Varykhalov:2010a,Walter:2011fj}) that was explained via reduction of the local symmetry for two carbon atoms in the graphene unit cell as a result of hybridization of Cu\,$3d$ states with different local character of the same interface Cu atom with $p_z$ orbitals of different carbon atoms in the unit cell~\cite{Vita:2014aa}.    

Beyond 3$d$ metals, intercalation of Eu and Cs between graphene and Ir(111) has been studied recently, revealing interesting details with respect to the interfacial process~\cite{Schumacher:2013ge,Schumacher:2013hl,Petrovic:2013vz}. For the initial intercalation of Eu a formation of compact islands and stripes quantized in size by the moir\'e unit cell are observed~\cite{Schumacher:2013ge}, supporting the previous reports on the growth of Ni underneath graphene on Rh(111)~\cite{Sicot:2012}.  Upon intercalation of larger amounts of Eu the structures grow in size yielding a pattern consisting of large islands and stripes divided by narrow channels. This growth behaviour is attributed both to the modulation of bonding strength between graphene and Ir(111) within the moir\'e unit cell and to the strain in the graphene layer generated during cooling due to the difference in the thermal expansion coefficients of graphene and Ir. At the atomic level Eu forms a $p(2\times2)$ superstructure with respect to graphene, being the lowest energy configuration for graphene/Eu/Ir(111) as confirmed by the DFT calculations. Further photoemission studies showed, that partial intercalation of Eu or Cs leads to a strong spatial modulation of the graphene doping level, graphene becomes $n$-doped above the intercalated patches whereas the nonintercalated regions remain $p$-doped~\cite{Schumacher:2013hl}. 

Metals being the most popular intercalation materials are complimented by semiconductors, such as Si~\cite{Meng:2012ee,Lizzit:2012hh}, or molecular species, including hydrogen~\cite{Riedl:2010du}, water~\cite{Feng:2012il,Feng:2013bp}, oxygen~\cite{Granas:2012cf,Larciprete:2012aaa,Jang:2013cn}, carbon monoxide~\cite{Granas:2013tl} or even C$_{60}$~\cite{Shikin:2000a,Varykhalov:2010}. Intercalation of these materials mostly aims at the electronic decoupling of the graphene layer form the underlying metal support. Silicon intercalated between graphene and Ru(0001) has been shown to yield well-ordered 2D patches and the disappearance of the strongly corrugated moir\'e structure~\cite{Cui:2012ki}. The latter observation coupled to the results of the corresponding photoemission studies were considered as indicative for the efficient decoupling of the graphene layer from the Ru substrate. This method has recently been further expanded by the oxidation of the interfacial metal silicide layer, resulting in an insulating SiO$_2$ film separating graphene from the metal as confirmed by photoemission~\cite{Lizzit:2012hh}. However, the investigations of the local crystallographic structure and electronic structure properties of this system are still missing, thus making it difficult to predict if this method can be used to efficiently decouple graphene nanostructures from the metal substrates.

The effective decoupling of graphene from metallic support can be achieved via intercalation of oxygen or CO~\cite{Zhang:2009qqq,Granas:2012cf,Larciprete:2012aaa,Granas:2013tl,Jang:2013cn}. These species intercalate in graphene/Ir(111) either in the form of atomic oxygen forming $p(2\times1)$-O layer~\cite{Granas:2012cf} or CO-molecules with $(3\sqrt{3}\times3\sqrt{3})R30^\circ$ structure~\cite{Granas:2013tl} underneath graphene. Graphene in these systems is fully decoupled and it is $p$-doped with the position of the Dirac point at $E-E_F=0.64$\,eV and $E-E_F=0.6$\,eV, respectively. For graphene/Ru(0001) the results are contradictory~\cite{Zhang:2009a,Sutter:2010a,Jang:2013cn}. These works show that oxygen can be placed underneath graphene as conformed by STM and thermal-programmed-desorption experiments and forms $(2\times2)$ or $(2\times1)$ superstructure between graphene and Ru(0001). However, STS experiments show that graphene in this system is $n$-doped with the position of the Dirac point at $E-E_F=-0.48$\,eV~\cite{Jang:2013cn}, that is opposite to the ARPES results where graphene was found highly $p$-doped with a Dirac point by $\approx0.5$\,eV higher $E_F$~\cite{Sutter:2010a}.

The interesting result was demonstrated in Ref.~\cite{Shikin:2000a} where intercalation of C$_{60}$ molecules in graphene/Ni(111) was assumed on the basis of spectroscopic data (photoelectron spectroscopy and electron energy loss spectroscopy). Later these results were examined with STM~\cite{Varykhalov:2010}. It was found that annealing of the thick layer of C$_{60}$ molecules leads to the desorption of most molecules, but also to its intercalation underneath graphene. The bias-dependent STM allows to discriminate between places of the graphene/C$_{60}$/Ni(111) structure where intercalation took place and not. These measurements unambiguously show that C$_{60}$ intercalates via certain interfacial channels, which can be induced by strain relaxation at defect sites of graphene/Ni(111).

\begin{figure}[t]
\includegraphics*[width=\linewidth]{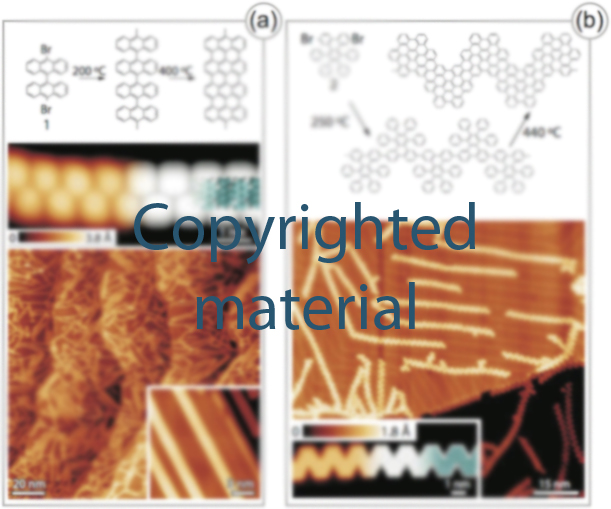}
\caption{(a) From top to bottom: (i) Reaction scheme from precursor 1 to straight $N=7$ GNRs; (ii) STM image taken after surface-assisted C-C coupling at $200^\circ$\,C but before the final cyclodehydrogenation step, showing a polyanthrylene chain (left) ($U_T=1.9\mathrm{V}$, $I_T=0.08\mathrm{nA}$), and DFT-based simulation of the STM image (right) with partially overlaid model of the polymer (blue, carbon; white, hydrogen); (iii)  Overview STM image after cyclodehydrogenation at $400^\circ$\,C, showing straight $N=7$ GNRs (RT) ($U_T=-3\mathrm{V}$, $I_T=0.03\mathrm{nA}$). The inset shows a higher-resolution STM image taken at $35$\,K ($U_T=-1.5\mathrm{V}$, $I_T=0.5\mathrm{nA}$). (b) Top: Reaction scheme from 6,11-dibromo-1,2,3,4-tetraphenyltriphenylene monomer 2 to chevron-type GNRs; Bottom: Overview STM image of chevron-type GNRs fabricated on a Au(111) surface ($T=35$\,K) ($U_T=-2\mathrm{V}$, $I_T=0.02\mathrm{nA}$). The inset shows a high-resolution STM image ($T=77$\,K) ($U_T=-2\mathrm{V}$, $I_T=0.5\mathrm{nA}$) and a DFT-based simulation of the STM image (greyscale) with partly overlaid molecular model of the ribbon (blue, carbon; white, hydrogen). Images are reproduced from Ref.~\cite{Cai:2010} with permission.}
\label{STM_GNR_Au111_Cai}
\end{figure}

\section{Graphene nanoribbons: synthesis and SPM studies}

\begin{figure*}[t]
\includegraphics*[width=\linewidth]{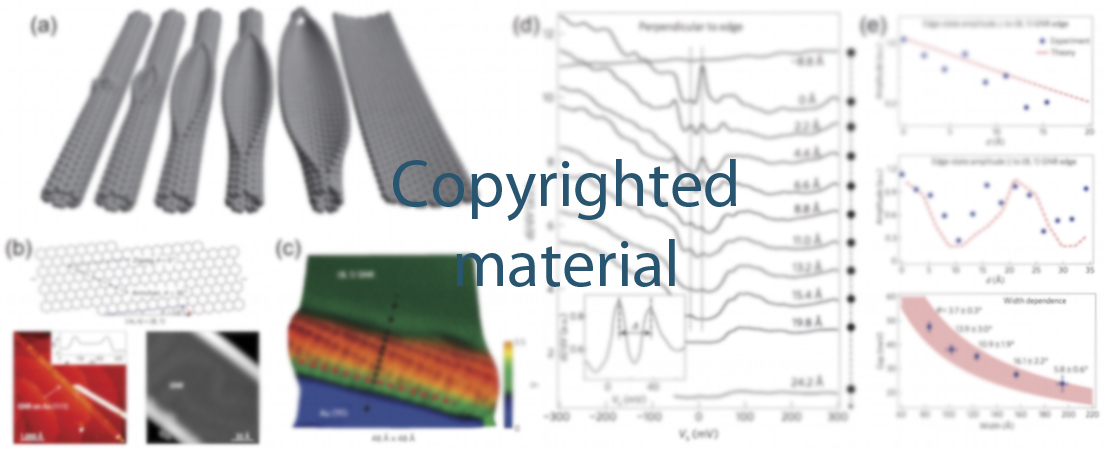}
\caption{(a) Representation of the gradual unzipping of one wall of a carbon nanotube to form a nanoribbon. Oxygenated sites are not shown. Represented from Ref.~\cite{Kosynkin:2009bw} with permission. (b) A schematic drawing of an $(8,1)$ GNR. The chiral vector $(n,m)$ connecting crystallographically equivalent sites along the edge defines the edge orientation of the GNR (black arrow). The blue and red arrows are the projections of the $(8,1)$ vector onto the basis vectors of the graphene lattice. Zigzag and armchair edges have corresponding chiral angles of $\theta=0^\circ$ and $\theta=30^\circ$, respectively, whereas the $(8,1)$ edge has an chiral angle of $\theta=5.8^\circ$. Lower part shows STM images of a monolayer GNR on Au(111) at room temperature (left) ($U_T=1.5\mathrm{V}$, $I_T=100\mathrm{pA}$) and higher resolution STM image of a GNR at $T=7$\,K (right) ($U_T=200\mathrm{mV}$, $I_T=30\mathrm{pA}$). (c) Atomically-resolved STM of the terminal edge of an $(8,1)$ GNR ($U_T=300\mathrm{mV}$, $I_T=60\mathrm{pA}$). (d) $dI/dV$ spectra measured at different positions across the edge of GNR as shown in (c). (e) From top to bottom: edge state amplitude perpendicular and parallel to the edge of an $(8,1)$ GNR and gap dependence on the width of GNRs. Data in (b-e) are reproduced from Ref.~\cite{Tao:2011kl} with permission.}
\label{STM_GNR_Au111_Tao}
\end{figure*}

A graphene nanoribbon (GNR) is a narrow strip of graphene, which morphology is determined by edge morphology and the width. In analogy to carbon nanotubes, the morphology of GNRs is characterized by the chiral index $(n, m)$ defining the edge translation vector (or, equivalently, by chiral angle, $\theta$) and classified as either armchair, zigzag or chiral [see Fig.~\ref{STM_GNR_Au111_Tao}(b)].

The GNR electronic band structure differs from that of graphene due to the quantum confinement effect perpendicular to the GNR axis. Depending on the edge morphology and the width, GNRs may be metallic, semimetallic or semiconducting~\cite{Nakada:1996us}. Even from the early theoretical studies employing tight binding approach it is known, that  zigzag nanoribbons (ZGNRs) display a sharp peak in the electronic density of states at the Fermi level, which is caused by a flat band characteristic of the zigzag edge of graphene and leads to a net spin polarization of the edge~\cite{Nakada:1996us,Fujita:1996vs,Wakabayashi:1999ti}. A small fundamental band gap opens up due to the antiferromagnetic coupling between the two spin-polarized edges. Calculations using the H\"uckel molecular orbital method~\cite{Nakada:1996us,Fujita:1996vs,Wakabayashi:1999ti,Ezawa:2006ic} or a two-dimensional free massless particle Dirac's equation with an effective speed of light ($\sim 10^6$\,m/s)~\cite{Brey:2006cb,Sasaki:2006wr,Abanin:2006ds} suggest that armchair nanoribbons (\mbox{AGNRs}) can be either semimetallic or semiconducting. An N-AGNR with N linear rows of atoms, is semimetallic if $N = 3p + 2$ ($p$ is an integer), and semiconducting otherwise. GNRs with Klein edges are always semiconducting. These observations are in good agreement with the data obtained with the first-principle approaches~\cite{Lee:2005ek,Son:2006ky}.

The above theoretical predictions made certain GNR types even more attractive for applications in electronics, than the parent gapless material. This attraction is warmed up by the fact that the modern experimental techniques allow to create GNRs with nearly atomic precision~\cite{Cai:2010}. The problem is that the major part of theoretical predictions was done without consideration of an additional but crucial component of any working device, which is the metallic contacts. We have shown above, that due to the non-negligable degree of hybridisation between metal and graphene valence band states, the linear dispersion of the $\pi$ states in the vicinity of $E_F$ characteristic for the free-standing graphene can be modified up to complete rearrangement of bands. Similar effects can also be anticipated in GNRs. Thus, on the one hand, theoretical consideration of GNRs have to be revisited in order to elucidate the role of the substrate. On the other hand, clearly, the results of calculations strongly depend on the technique of calculating the electronic structure and such studies have to be strongly linked to the experiments.

Recently, several ways were proposed for the preparation of GNRs. Among them are lithographic patterning (limited by the method resolution of $\sim20$\,nm)~\cite{Han:2007bl}, chemical sonication~\cite{Li:2008ht}, gas-etching chemistry of GNRs produced by e-beam lithography (nanoribbons can be narrowed to the $\sim5$\,nm width)~\cite{Wang:2010cf}, cutting of a graphene layer along crystallographic axes by thermally activated metallic clusters of Fe or Ni~\cite{Datta:2008if,Ci:2008je,Campos:2009cd}, \textit{etc}. However all these methods have either limited spacial resolution that does not allow to produce narrow nanoribbons or produce GNRs with edges of uncontrollable morphology that can lead to the uncertainty in the electronic transport properties of the devices built on their basis, because as was shown the geometry as well as defects of the GNRs edges have a very high impact on the electron and spin transport properties of nanoribbons~\cite{Wakabayashi:1999ti,Pisani:2007ge,Saffarzadeh:2011fd,Fu:2012de,Farghadan:2012gj,Hawkins:2012kv}. 

GNRs with atomic precision can be prepared in different ways. The first controllable synthesis was demonstrated in Ref.~\cite{Cai:2010}, where a bottom-up approach to grow subnanometer wide armchair GNRs on Au(111) with clean edges was demonstrated (Fig.~\ref{STM_GNR_Au111_Cai}). Formation of these structures is performed in two thermal activation steps shown in panels (a) and (b) for two different organic precursors. In the second step the initially formed linear polymer chains undergo a surface-assisted cyclodehydrogenation and an extended fully aromatic systems are formed. The experimental and simulated STM images of such GNRs are shown in Fig.~\ref{STM_GNR_Au111_Cai}. Subsequent STS measurements yield a band gap of $2.3\pm0.1$\,eV for 7-AGNR ($1.4\pm0.1$\,eV for 13-AGNRs~\cite{Chen:2013fa}) compared to $2.3-2.7$\,eV obtained from quasiparticle $GW$ calculation corrected for the image charge on the metallic substrate~\cite{Ruffieux:2012et}. Parallel ARPES and IPES experiments give for straight 7-AGNRs and 13-AGNRs values for the band gap of $2.8\pm0.4$\,eV and $1.6\pm0.4$\,eV, respectively. For the chevron-type GNRs [Fig.~\ref{STM_GNR_Au111_Cai}(b)], they give a band gap of $3.1\pm0.4$\,eV. Also these measurements showed that electrons in GNRs cannot be considered as massless Dirac fermions as the carriers exhibit now a finite effective mass $m^*=0.21m_0$, where $m_0$ is the free electron mass. For the ARPES experiments, which require the unidirectional alignment of GNRs at the macroscopic scale, the similar growth procedure was used on the stepped Au(788) surface which consists of the $\{111\}$ terraces of $3.83$\,nm width~\cite{Ruffieux:2012et,Linden:2012fu}. Aligned assemblies of both types, straight and chevron-type, GNRs can be grown in such case. Later, these results were examined within DFT and many-body electron approaches~\cite{Liang:2012wi}. These calculations indicate that electron transfer exists from any type of GNRs on Au(111) that leads to the surface polarization, that is responsible for the width of the band gap of GNRs. The calculated energy gaps are $2.85$\,eV and $2.96$\,eV for 7-AGNRs and chevron-type GNR, respectively, that agrees well with the experimental data.

\begin{figure}[t]
\includegraphics*[width=0.95\linewidth]{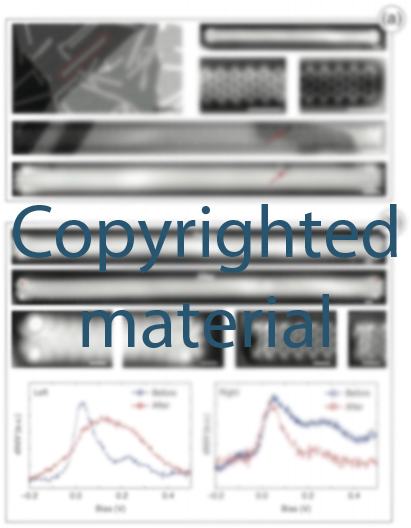}
\caption{(a) Clockwise: STM overview image showing a free GNRs ($U_T=50\mathrm{mV}$, $I_T=5\mathrm{pA}$); Free GNR imaged with a CO-terminated tip ($U_T=10\mathrm{mV}$, $I_T=5\mathrm{pA}$); Constant-height high-resolution nc-AFM images of the middle and the zigzag end of a GNR obtained with a CO-terminated tip (AFM set-point offset by $30$\,pm); Constant-height nc-AFM/STM images of 22 monomer unit long GNR with a single missing benzene ring marked by the red arrow (AFM set-point offset by $48$\,pm). In the first image scale bar is $10$\,nm, other scale bars are $1$\,nm. (b) Controlled atomic-scale modification of the GNR reduces vibronic coupling: High-resolution STM image of a free GNR before ($U_T=50\mathrm{mV}$, $I_T=5\mathrm{pA}$) and after ($U_T=50\mathrm{mV}$, $I_T=20\mathrm{pA}$) a bias pulse has been used to modify the left end of the GNR. Middle row, from left to right: Zoomed-in STM images of the GNR ends after the modification ($U_T=50\mathrm{mV}$, $I_T=50\mathrm{pA}$; $U_T=10\mathrm{mV}$, $I_T=2\mathrm{pA}$); Atomically resolved AFM image of the contacted GNR (AFM set-point offset by $120$\,pm); AFM image of the same ribbon as in previous image, but with the tip 80 pm closer to the sample. Scale bar is $0.5$\,nm is all images. Lower row: $dI/dV$ spectra recorded at the left and right ends of the GNR before (blue) and after (red) the modification. All the images and spectra have been acquired with a CO-terminated tip. All data are reproduced from Ref.~\cite{vanderLit:2013jf} with permission.}
\label{STMAFM_GNR_Au111_vanderLit}
\end{figure}

Another approach to obtain GNRs, the so-called unzipping of carbon nanotubes, allows flexible variation of GNR width, length, chirality, and substrate~
\cite{Kosynkin:2009bw,Jiao:2010de} [Fig.~\ref{STM_GNR_Au111_Tao}(a)]. In this method the nanotubes were placed in an organic solution in the ultrasonic bath and then nanoribbons were deposited on a Au(111) substrate and then their structure and electronic properties were studied with STM/STS~\cite{Tao:2011kl} [Fig.~\ref{STM_GNR_Au111_Tao}(b)]. This method allows to prepare GNRs of different chiralities ($3.7^\circ<\theta<16.1^\circ$) and width ($> 500$\,nm) on Au(111). As was found in STM the edge of GNRs is ``bumped'' indicating the effect of the GNR-metal interaction [Fig.~\ref{STM_GNR_Au111_Tao}(c)]. Spectroscopic studies (STS) across the edge of GNR [(c) and (d)] show the existence of two peaks located around $E_F$. The measured edge-state energy splitting shows a clear inverse correlation with GNR width [Fig.~\ref{STM_GNR_Au111_Tao}(e)] and the gap values tend to be smaller than those observed for lithographically patterned GNRs (probably because of uncertainty in the edge structure of lithographically obtained GNRs~\cite{Han:2007bl}). The observed spectroscopic feature (i.e. a double-peak in the local density of states near $E_F$  close to the edge with a peak distance scaling inversely with the width of the GNR) was ascribed to an antiferromagnetic coupling between opposite edges~\cite{Feldner:2011cm}, as further evidenced by comparison with results from a Hubbard model Hamiltonian~\cite{Tao:2011kl} [hopping integral, $t=2.7$\,eV; on-site Coulomb repulsion, $U=0.5t=1.35$\,eV]. These theoretical studies were extended for the wider range of $\theta$ and in the case of $\theta =30^\circ$, i.e. AGNR, the edge states were not present. The latter is in agreement with the results on AGNR/metal systems obtained by means of plane-wave DFT including semi-empirical dispersion correction~\cite{Steinkasserer:2014iv}.

In agreement with the results by Tao \textit{et al.}~\cite{Tao:2011kl}, other experimental studies provide direct or indirect evidence for the presence of the edge states in supported GNRs~\cite{Ritter:2009hw,Pan:2012hp}. Such edge states might be exploited for a multitude of spintronic applications, although it is not fully clear if the edge states contribute to the measured transport properties of nano structures at all. To elucidate the role of the substrate and of edge termination, ZGNRs close to realistic width on (111) surfaces of Cu, Ag and Au were studied recently by means of DFT-D2~\cite{Li:2013ih}. Although it was found that GNRs possess edge states, independent on whether they are H-free or H-terminated, they do not exhibit a significant magnetisation at the edge. These results are explained by the different interaction and charge transfer between the GNRs and the substrates and show that edge magnetism in zigzag GNRs can be destroyed even upon deposition on a substrate which interacts weakly with graphene. Only in the case of H-terminated GNRs on Au(111) the interaction at the edge was found to be sufficiently weak to not affect the electronic and magnetic properties of the edge states significantly.

\begin{figure*}[th]
\includegraphics*[width=\textwidth]{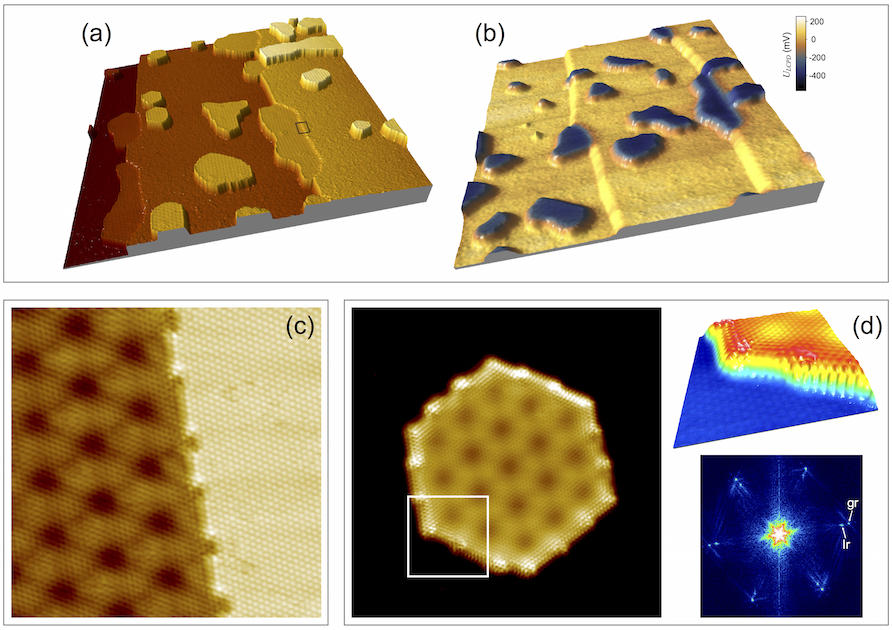}
\caption{(a) and (b) 3D views of the large scale STM ($160\times160\mathrm{nm}^2$, $U_T=300\mathrm{mV}$, $I_T=0.8\mathrm{nA}$) and AFM images ($150\times150\mathrm{nm}^2$, $\Delta f=-670\mathrm{mHz}$), respectively, of graphene flakes and QDs on Ir(111). In (b) the topography image is overlaid with the results of KPFM imaging obtained simultaneously during AFM scanning. (c) Atomically resolved border between graphene flake and Ir(111) marked in (a) by black rectangle ($10\times10\mathrm{nm}^2$, $U_T=-50\mathrm{mV}$, $I_T=30\mathrm{nA}$). (d) STM image of the single zigzag-edge terminated GQD on Ir(111) ($20.2\times20.2\mathrm{nm}^2$, $U_T=300\mathrm{mV}$, $I_T=1.5\mathrm{nA}$). Right-hand side shows the zoomed area marked by the white rectangle ($4.1\times4.1\mathrm{nm}^2$, $U_T=300\mathrm{mV}$, $I_T=1.5\mathrm{nA}$) and the corresponding FFT spectrum of the STM image of GND/Ir(111).}
\label{STM_GQD_Ir111_Dedkov}
\end{figure*}

The interaction of GNRs with Au(111) and effect of the edge-termination was studied by combination of STM and AFM in Ref.~\cite{vanderLit:2013jf} [Fig.~\ref{STMAFM_GNR_Au111_vanderLit}]. In this work GNRs were prepared by the method described above~\cite{Cai:2010,Ruffieux:2012et}. The simultaneous STM and AFM imaging of GNR with the tip terminated by CO allows to clearly identify all small changes in the structure of the nanoribbons. For example, the missing benzene ring marked in Fig.~\ref{STMAFM_GNR_Au111_vanderLit}(a) by the red arrow is clearly resolved in AFM, but not very well resolved in STM. Later in the same work, one of the edges of the GNR was modified via removal of one of the H-atom that leads to the formation of the bonds between edge of GNR and underlying Au(111). This place is now imaged as a dark (less repulsive) area at the edge of GNR [Fig.~\ref{STMAFM_GNR_Au111_vanderLit}(b)]. It is interesting to note that after such modification the GNRs cannot be moved by the STM tip (as was possible before modification), indicating the stronger interaction between GNR and substrate. Formation of the bond also changes the electronic spectrum of GNR and leads to the suppression of the vibronic tunnelling modes and STS spectrum shows now a broad peak for the left edge (H-atom is missing) of GNR [Fig.~\ref{STMAFM_GNR_Au111_vanderLit}(b)]. Also the STS spectrum obtained on the intact edge (several tens nm away from the modified edge) demonstrates the reduction of the intensity of the vibronic mode although the elastic peak remains unchanged. As was found in these experiments the bulk electronic structure of GNR is unchanged during all modifications, but the formed contacts at the edges might modify the transport properties of GNR as the vibronic modes in the tunnelling conductivity are strongly altered.

While the interaction of nanoribbons with a noble metal contact is rather weak, their adsorption on metals like Pd and Ti, which are widely used in graphene-based devices, involve much stronger state mixing. The latter systems were considered in Ref.~\cite{Archambault:2013fd} by means of DFT.  These calculations show that the so-called metal-induced gap states appear with a high density in the energy gap of GNRs at the metal-GNR interface contact and that they effectively penetrate for the large distance in GNR. These states can even lead to the shortening of the metal contacts through GNR that limits the application of the ultimately small GNR terminals. This work showed the importance of the consideration of the metal-GNR interfaces where intermixing of the valence band states at the interface of both electrodes is expected.

\section{Graphene nanoflakes and quantum dots: synthesis and SPM studies}

Graphene quantum dots (GQDs) are small graphene fragments (typically with the lateral size below $20$\,nm), where electronic transport is confined in all three spatial dimensions. GQDs can be fabricated by fragmentation or ``cutting'' of graphene sheets (top-down approach). Alternatively, large graphene-like molecules can be synthesized with well-defined molecular structure (bottom-up approach). 

The early work on GQDs was dominated by the investigation of their physical properties. More recently, GQDs have been chemically modified and used for the first time in applications in the area of energy conversion, bioanalysis and sensors. Another exciting perspective is to use GQDs as spin qubits~\cite{Trauzettel:2007}. The basic prerequisite is a very long spin-coherence time, which might exist in graphene due to the absence of hyperfine coupling in isotopically pure material and the small spin-orbit coupling. However, since graphene provides no natural gap, it is difficult to control the electron number. Moreover, the 2D sub-lattice symmetry makes the QD properties very susceptible to the atomic edge configuration~\cite{Trauzettel:2007,DasSarma:2011br} unlike conventional QDs. As a result, chaotic Dirac billiards have been predicted~\cite{Rycerz:2007} and were even claimed to be realized~\cite{Ponomarenko:2008ej}; i.\,e., the wave functions are assumed to be rather disordered. To achieve improved control of GQDs, the QD edges must be well defined.

Graphene flakes or quantum dots on metals were successfully synthesised at high temperature via decomposition of preadsorbed organic molecules, like hydrocarbons (C$_2$H$_4$~\cite{Hamalainen:2011ja,Subramaniam:2012fp} or coronene~\cite{Coraux:2009}), or unfolding of redeposited C$_{60}$ molecules~\cite{Lu:2011bg}. Fig.~\ref{STM_GQD_Ir111_Dedkov} shows results of (a) STM and (b) combined AFM/KFM imaging of graphene flakes and QD synthesised on Ir(111) from C$_2$H$_4$ by means of thermal programming growth (TPG) procedure~\cite{Dedkov:2014di}. The next panels in the same figure show the atomically-resolved STM images (c) of the border between graphene flake and Ir(111) marked by the black frame in (a) and (d) single graphene QD and the corresponding edge marked by the white frame and the respective FFT spectrum of the STM image of QD where the spots corresponding to the reciprocal lattices of Ir and graphene are marked. As can be deduced from panels (a,c,d), the graphene flakes and QD have well-ordered structure terminated by the zigzag edges which are strongly bonded to the Ir(111) surface. The strong surface inhomogeneity of the graphene-QD/Ir(111) system allows to perform accurate AFM/KPFM measurements of $U_{LCPD}$. Fig.~\ref{STM_GQD_Ir111_Dedkov}(b) shows the 3D representation of topography of the graphene-QD/Ir(111) system measured in AFM mode overlaid with the simultaneously measured KPFM signal. The measured difference of $U_{LCPD}$ is $\approx600$\,meV that is smaller than the difference of work functions of Ir(111) and graphene extracted from STS ($1.1\pm0.3$\,eV)~\cite{Forster:2012aa} and LEEM ($1.6$\,eV)~\cite{Starodub:2012cb}. This effect can be connected with the smearing effect of the macroscopic scanning tip with a sharp apex~\cite{Jacobs:1998vo,Baier:2012ey}.

Direct insight into electronic structure and properties of QD is provided by STS which maps out the LDOS of confined charge carriers and, at the same time, determines the shape of the QD atom by atom. Recently, several groups investigated \textit{in situ} prepared GQDs with exclusive zigzag edges, which are supported by Ir(111)~\cite{Hamalainen:2011ja,Phark:2011de,Subramaniam:2012fp,Altenburg:2012kp}. The presence of edge-states was not detected~\cite{Li:2013ji}. The experimental results are supported by DFT calculations. These observations are explained by a hybridisation of the GQDs $p_z$ orbitals and the substrate valence band states (here: Ir $5d_{z^2}$ surface state lying at $E_F$), which gradually decreases in strength from the edge towards the center of the GQD. However, these works, presenting the similar experimental data on the confinement of the electronic states in GQDs, do not provide a clear answer to what extent the graphene electronic states contribute to the tunnelling current compared to the contribution from the substrate. 

\begin{figure*}[t]
\includegraphics*[width=\textwidth]{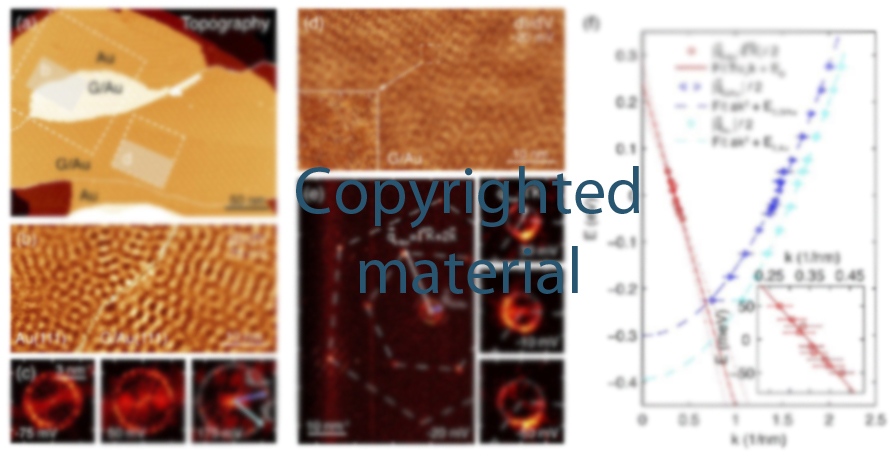}
\caption{(a) Topography of an approximately $400\times160$\,nm$^2$ large flake used for $dI/dV$ mappings. (b) $dI/dV$ map across the rim of a large graphene flake showing the quasiparticle interferences of the Au surface state ($U_T=-75\mathrm{mV}$, $I_T=1\mathrm{nA}$). (c) Fast Fourier transform of a $109\times109$\,nm$^2$ large area partially depicted in (b) at selected bias voltages. (d) $dI/dV$ map on the graphene flake with atomic resolution ($U_T=-20\mathrm{mV}$, $I_T=1\mathrm{nA}$). (e) Fast Fourier transform of a $54\times54$\,nm$^2$ map partially depicted in (d) showing a rich structure including atomic, moir\'e, and herringbone features as well as intervalley scattering features. The right column shows magnifications of the intervalley scattering rings ($3.5\times3.5$\,nm$^{-2}$) at different bias voltages. (f) The Au(111) surface state dispersion for pristine Au(111) (cyan diamonds) as well as for graphene/Au(111) (blue triangles). Dispersion relation of the graphene electrons determined from the intervalley scattering (red squares) and corresponding fit (red line) including uncertainty (red dotted lines) determined from a series of atomically resolved constant-energy mappings. The $k$ values are plotted with respect to the $\Gamma$-point in the case of the surface state and with respect to the $K$-point in case of graphene. Data are reproduced from Ref.~\cite{Leicht:2014jy} with permission.}
\label{STM_GQD_Au111_Leicht}
\end{figure*}

This problem was recently addressed in Ref.~\cite{Leicht:2014jy}, whose main results are presented in Fig.~\ref{STM_GQD_Au111_Leicht}. In this work graphene flakes were initially produced on Ir(111) via TPG procedure and then they were decoupled from it via intercalation of thick layer of Au. In this case the graphene-flakes/Au(111) system of very high quality is formed demonstrating herringbone reconstruction of Au(111) as well as a moir\'e structure on the graphene flakes due to the lattice mismatch of graphene and Au(111). Compared to graphene/Ir(111) where graphene flakes are bonded to the Ir substrate at the edges~\cite{Lacovig:2009}, they are very mobile on the Au(111) surface. This is conformed by the possibility to move graphene flakes across the surface by the STM tip with only moderate tunnelling currents (typical parameters are $U_T=1$\,V and $I_T=1$\,nA).

The weak interaction between the graphene flakes and Au(111) allows to carefully follow the dispersion of the electronic states obtained from STS experiments (see Fig.~\ref{STM_GQD_Au111_Leicht}). Panel (a) shows an STM image of a single graphene flake on Au(111) and panel (b) presents the corresponding $dI/dV$ map measured at $U_T=-75$\,mV of the area from (a) where simultaneously the Au(111) surface and the graphene-flake/Au(111) interface are probed. One can clearly see that on both sides of the border the electronic standing waves have the same character that indicates the same origin of the electronic states -- surface sate of Au(111). Similar spectroscopy maps were acquired at different bias voltages. FFT analysis of these $dI/dV$ maps demonstrates the existence of two circles in the reciprocal space with the radii of $q_{gr/Au}$ and $q_{Au}$, which can be assigned to the surface state of Au(111) on graphene-covered and graphene-free surfaces. The resulting FFT data allow to plot the respective dispersions of the electronic states $E(k)$, where $q_{Au,gr/Au}=2k$ [Fig.~\ref{STM_GQD_Au111_Leicht}(f)]. The extracted effective mass is similar for both electronic states, $m^*=0.26m_e$.

The similar $dI/dV$ measurements with atomic resolution were performed on the same graphene flake [Fig.~\ref{STM_GQD_Au111_Leicht}(d)]. The extracted FFT pictures show additional features: (i) six spots corresponding to the reciprocal lattice of graphene, which are superposed by the spots originating from the reconstruction of Au(111) and the moir\'e structure (large hexagon); (ii) ring-like features which build a hexagon corresponding to the $(\sqrt{3}\times\sqrt{3})R30^\circ$ structure in the real space and related to the intervalley scattering ($\vec{q}_{\mathrm{inter}}=\Gamma K-2\vec{k}$). The obtained FFT data at different bias voltages allow to plot the dispersion of these electronic states [red squares in Fig.~\ref{STM_GQD_Au111_Leicht}(f); they are plotted with respect to the $\mathrm{K}$ point of the graphene Brillouine zone]. The linear fit of these data gives the Fermi velocity of $v_F=(1.1\pm0.2)\cdot10^6$\,m/s and the position of the Dirac point at $E-E_F=0.24\pm0.05$\,eV, i.\,e. graphene is $p$-doped as found in the earlier photoemission studies of the graphene/Au(111) system~\cite{Enderlein:2010}. These measurements performed on the same graphene flake show the spectroscopic features from the surface state of Au(111) and from the intervalley scattering of the graphene Dirac fermions allow to give an answer about different contributions in the imaging of graphene on metals~\cite{Hamalainen:2011ja,Phark:2011de,Subramaniam:2012fp,Altenburg:2012kp,Leicht:2014jy,Jolie:2014ev,GarciaLekue:2014ce}.

The similar experiments were performed on the GQDs/oxygen/Ir(111) system~\cite{Jolie:2014ev}. A linear fit of the data gives the Fermi velocity of $v_F=(0.96\pm0.07)\cdot10^6$\,m/s and the position of the Dirac point at $E-E_F=0.64\pm0.07$\,eV, which agrees well with the data obtained in ARPES experiments in the same and similar works~\cite{Jolie:2014ev,Larciprete:2012aaa}.

Preparation of graphene flakes or GNDs on noble metals might lead to the formation of highly strained graphene nonobubbles, as it was demonstrated in Ref.~\cite{Levy:2010} for graphene-flakes/Pt(111). Such structures have unique electronic properties as was demonstrated in STS measurements. The STS $dI/dV$ spectra measured across such graphene nanobubble show several spectral features, which can be assigned to the Landau levels in graphene. The nature of this effect is the existence of the so-called pseudo magnetic field (of more than $300$\,T) originating from the strain appearing in the graphene nanobubbles during preparation (different expansion coefficients of graphene and Pt). The similar STS signatures were found for graphene nanobubbles formed by the intercalation of oxygen in graphene/Ru(0001)~\cite{Lu:2014dn}, finally giving a chance to form the regular arrays of the strain-engineered graphene-based structures.

\section{Conclusions}

The present review demonstrates the powerful potential of the modern SPM methods in conjunction with the corresponding spectroscopy for the investigation of the graphene-based systems on metallic surfaces. These recent experimental works are devoted to the studies of the graphene-metal interfaces and nanostructures where different phenomena were observed: localisation of the electronic states in graphene nanoribbons and GND, edge electron scattering in different geometries, demonstration of the Landau levels in strained graphene nanostructures, etc. For simple graphene-flakes on metals it was shown that it is possible to discriminate between graphene- and metal-related contributions in the electronic structure of the system at the nanoscale. All these results are found to be in very good agreement with state-of-the-art DFT calculations that allow to model more complicated graphene-based nanostructures. Such an approach permits to study the more realistic graphene-metal systems which will find a broad utilisation in future electron- and spin-transport applications and devices.

\begin{acknowledgement}
We would like to acknowledge our colleagues for the useful discussions, in particular, K. Horn, S. Torbr\"ugge, T. H\"anke, M. Breitschaft, M. Breusing, R. Kremzow, N. Berdunov, Th. K\"onig, Th. Kampen, O. Schaff, A. Thissen, P. Leicht, J. Tesch, L. Zielke. E.N.V. and M.F. appreciate the support from the German Research Foundation (DFG) through the Priority Programme (SPP) 1459 ``Graphene'' and European Science Foundation (ESF) via the EUROCORES Programme EuroGRAPHENE (Project ``SpinGraph'').
\end{acknowledgement}

\bibliographystyle{pss}


\begin{thebibliography}{[100]}

\bibitem{Novoselov:2004a}
 \textsc{K.~Novoselov},  \textsc{A.~Geim},  \textsc{S.~Morozov},
  \textsc{D.~Jiang},  \textsc{Y.~Zhang},  \textsc{S.~Dubonos},
  \textsc{I.~Grigorieva},  and  \textsc{A.~Firsov},
 \jr{Science} \textbf{306}, 666 (2004).


\bibitem{Novoselov:2005}
 \textsc{K.~Novoselov},  \textsc{A.~Geim},  \textsc{S.~Morozov},
  \textsc{D.~Jiang},  \textsc{M.~Katsnelson},  \textsc{I.~Grigorieva},
  \textsc{S.~Dubonos},  and  \textsc{A.~Firsov},
 \jr{Nature} \textbf{438}, 197 (2005).


\bibitem{Zhang:2005}
 \textsc{Y.~Zhang},  \textsc{Y.~Tan},  \textsc{H.~Stormer},  and
  \textsc{P.~Kim},
 \jr{Nature} \textbf{438}, 201 (2005).


\bibitem{CastroNeto:2009}
 \textsc{A.~Castro~Neto},  \textsc{F.~Guinea},  \textsc{N.~Peres},
  \textsc{K.~Novoselov},  and  \textsc{A.~Geim},
 \jr{Rev. Mod. Phys.} \textbf{81}, 109 (2009).


\bibitem{Geim:2009}
 \textsc{A.~Geim},
 \jr{Science} \textbf{324}, 1530 (2009).


\bibitem{Hagstrom:1965vh}
 \textsc{S.~Hagstrom},  \textsc{H.\,B. Lyon},  and  \textsc{G.\,A.
  Somorjai},
 \jr{Phys. Rev. Lett.} \textbf{15}, 491 (1965).


\bibitem{May:1969uj}
 \textsc{J.\,W. May},
 \jr{Surf. Sci.} \textbf{17}, 267 (1969).


\bibitem{Grant:1970}
 \textsc{J.\,T. Grant} and  \textsc{A.~Haas},
 \jr{Surf. Sci.} \textbf{21}, 76 (1970).


\bibitem{Land:1992}
 \textsc{T.\,A. Land},  \textsc{T.~Michely},  \textsc{R.~Behm},  \textsc{J.\,C.
  Hemminger},  and  \textsc{G.~Comsa},
 \jr{Surf. Sci.} \textbf{264}, 261 (1992).


\bibitem{Li:2009}
 \textsc{X.~Li},  \textsc{W.~Cai},  \textsc{J.~An},  \textsc{S.~Kim},
  \textsc{J.~Nah},  \textsc{D.~Yang},  \textsc{R.~Piner},
  \textsc{A.~Velamakanni},  \textsc{I.~Jung},  \textsc{E.~Tutuc},
  \textsc{S.\,K. Banerjee},  \textsc{L.~Colombo},  and  \textsc{R.\,S.
  Ruoff},
 \jr{Science} \textbf{324}, 1312 (2009).


\bibitem{Bae:2010}
 \textsc{S.~Bae},  \textsc{H.~Kim},  \textsc{Y.~Lee},  \textsc{X.~Xu},
  \textsc{J.\,S. Park},  \textsc{Y.~Zheng},  \textsc{J.~Balakrishnan},
  \textsc{T.~Lei},  \textsc{H.\,R. Kim},  \textsc{Y.\,I. Song},  \textsc{Y.\,J.
  Kim},  \textsc{K.\,S. Kim},  \textsc{B.~Ozyilmaz},  \textsc{J.\,H. Ahn},
  \textsc{B.\,H. Hong},  and  \textsc{S.~Iijima},
 \jr{Nature Nanotech.} \textbf{5}, 574 (2010).


\bibitem{Tao:2012tc}
 \textsc{L.~Tao},  \textsc{J.~Lee},  \textsc{M.~Holt},  \textsc{H.~Chou},
  \textsc{S.\,J. McDonnell},  \textsc{D.\,A. Ferrer},  \textsc{M.\,G. Babenco},
   \textsc{R.\,M. Wallace},  \textsc{S.\,K. Banerjee},  and  \textsc{R.\,S.
  Ruoff},
 \jr{J. Phys. Chem. C} \textbf{116}, 24068 (2012).


\bibitem{Wintterlin:2009}
 \textsc{J.~Wintterlin} and  \textsc{M.\,L. Bocquet},
 \jr{Surf. Sci.} \textbf{603}, 1841 (2009).


\bibitem{Voloshina:2012c}
 \textsc{E.~Voloshina} and  \textsc{Y.~Dedkov},
 \jr{Phys. Chem. Chem. Phys.} \textbf{14}, 13502 (2012).


\othercit
\bibitem{Dedkov:2012book}
 \textsc{Y.\,S. Dedkov},  \textsc{K.~Horn},  \textsc{A.~Preobrajenskij},  and
  \textsc{M.~Fonin},
{Epitaxial Graphene on Metals},
 in: Graphene Nanoelectronics, edited by H.~Raza,  (Springer, Berlin, July
  2012).


\bibitem{Voloshina:2014jl}
 \textsc{E.\,N. Voloshina} and  \textsc{Y.\,S. Dedkov},
 \jr{Materials Research Express} \textbf{1}, 035603 (2014).


\bibitem{Cai:2010}
 \textsc{J.~Cai},  \textsc{P.~Ruffieux},  \textsc{R.~Jaafar},
  \textsc{M.~Bieri},  \textsc{T.~Braun},  \textsc{S.~Blankenburg},
  \textsc{M.~Muoth},  \textsc{A.\,P. Seitsonen},  \textsc{M.~Saleh},
  \textsc{X.~Feng},  \textsc{K.~M{\"u}llen},  and  \textsc{R.~Fasel},
 \jr{Nature} \textbf{466}, 470 (2010).


\bibitem{Ruffieux:2012et}
 \textsc{P.~Ruffieux},  \textsc{J.~Cai},  \textsc{N.\,C. Plumb},
  \textsc{L.~Patthey},  \textsc{D.~Prezzi},  \textsc{A.~Ferretti},
  \textsc{E.~Molinari},  \textsc{X.~Feng},  \textsc{K.~M{\"u}llen},
  \textsc{C.\,A. Pignedoli},  and  \textsc{R.~Fasel},
 \jr{ACS Nano} \textbf{6}, 6930 (2012).


\bibitem{Linden:2012fu}
 \textsc{S.~Linden},  \textsc{D.~Zhong},  \textsc{A.~Timmer},
  \textsc{N.~Aghdassi},  \textsc{J.~Franke},  \textsc{H.~Zhang},
  \textsc{X.~Feng},  \textsc{K.~M{\"u}llen},  \textsc{H.~Fuchs},
  \textsc{L.~Chi},  and  \textsc{H.~Zacharias},
 \jr{Phys. Rev. Lett.} \textbf{108}, 216801 (2012).


\bibitem{Talirz:2013es}
 \textsc{L.~Talirz},  \textsc{H.~S{\"o}de},  \textsc{J.~Cai},
  \textsc{P.~Ruffieux},  \textsc{S.~Blankenburg},  \textsc{R.~Jafaar},
  \textsc{R.~Berger},  \textsc{X.~Feng},  \textsc{K.~M{\"u}llen},
  \textsc{D.~Passerone},  \textsc{R.~Fasel},  and  \textsc{C.\,A.
  Pignedoli},
 \jr{J. Am. Chem. Soc.} \textbf{135}, 2060 (2013).


\bibitem{vanderLit:2013jf}
 \textsc{J.~van\,der Lit},  \textsc{M.\,P. Boneschanscher},
  \textsc{D.~Vanmaekelbergh},  \textsc{M.~Ij{\"a}s},  \textsc{A.~Uppstu},
  \textsc{M.~Ervasti},  \textsc{A.~Harju},  \textsc{P.~Liljeroth},  and
  \textsc{I.~Swart},
 \jr{Nature Communications} \textbf{4}, 2023 (2013).


\bibitem{Zhang:2014em}
 \textsc{Y.~Zhang},  \textsc{Y.~Zhang},  \textsc{G.~Li},  \textsc{J.~Lu},
  \textsc{X.~Lin},  \textsc{S.~Du},  \textsc{R.~Berger},  \textsc{X.~Feng},
  \textsc{K.~M{\"u}llen},  and  \textsc{H.\,j. Gao},
 \jr{Appl. Phys. Lett.} \textbf{105}, 023101 (2014).


\bibitem{Subramaniam:2012fp}
 \textsc{D.~Subramaniam},  \textsc{F.~Libisch},  \textsc{Y.~Li},
  \textsc{C.~Pauly},  \textsc{V.~Geringer},  \textsc{R.~Reiter},
  \textsc{T.~Mashoff},  \textsc{M.~Liebmann},  \textsc{J.~Burgd{\"o}rfer},
  \textsc{C.~Busse},  \textsc{T.~Michely},  \textsc{R.~Mazzarello},
  \textsc{M.~Pratzer},  and  \textsc{M.~Morgenstern},
 \jr{Phys. Rev. Lett.} \textbf{108}, 046801 (2012).


\bibitem{Altenburg:2012kp}
 \textsc{S.\,J. Altenburg},  \textsc{J.~Kr{\"o}ger},  \textsc{T.\,O. Wehling},
  \textsc{B.~Sachs},  \textsc{A.\,I. Lichtenstein},  and
  \textsc{R.~Berndt},
 \jr{Phys. Rev. Lett.} \textbf{108}, 206805 (2012).


\bibitem{Li:2013ji}
 \textsc{Y.~Li},  \textsc{D.~Subramaniam},  \textsc{N.~Atodiresei},
  \textsc{P.~Lazic},  \textsc{V.~Caciuc},  \textsc{C.~Pauly},
  \textsc{A.~Georgi},  \textsc{C.~Busse},  \textsc{M.~Liebmann},
  \textsc{S.~Bl{\"u}gel},  \textsc{M.~Pratzer},  \textsc{M.~Morgenstern},  and
  \textsc{R.~Mazzarello},
 \jr{Adv. Mater.} \textbf{25}, 1967 (2013).


\bibitem{Leicht:2014jy}
 \textsc{P.~Leicht},  \textsc{L.~Zielke},  \textsc{S.~Bouvron},
  \textsc{R.~Moroni},  \textsc{E.~Voloshina},  \textsc{L.~Hammerschmidt},
  \textsc{Y.\,S. Dedkov},  and  \textsc{M.~Fonin},
 \jr{ACS Nano} \textbf{8}, 3735 (2014).


\bibitem{Giessibl:2003aaa}
 \textsc{F.\,J. Giessibl},
 \jr{Rev. Mod. Phys.} \textbf{75}, 949 (2003).


\bibitem{Giessibl:2011fw}
 \textsc{F.\,J. Giessibl},  \textsc{F.~Pielmeier},  \textsc{T.~Eguchi},
  \textsc{T.~An},  and  \textsc{Y.~Hasegawa},
 \jr{Phys. Rev. B} \textbf{84}, 125409 (2011).


\bibitem{Giessibl:2001ig}
 \textsc{F.\,J. Giessibl},
 \jr{Appl. Phys. Lett.} \textbf{78}, 123 (2001).


\bibitem{Sader:2004kt}
 \textsc{J.\,E. Sader} and  \textsc{S.\,P. Jarvis},
 \jr{Appl. Phys. Lett.} \textbf{84}, 1801 (2004).


\bibitem{Nonnenmacher:1991}
 \textsc{M.~Nonnenmacher},  \textsc{M.\,P. O'Boyle},  and  \textsc{H.\,K.
  Wickramasinghe},
 \jr{Appl. Phys. Lett.} \textbf{58}, 2921 (1991).


\bibitem{Melitz:2011}
 \textsc{W.~Melitz},  \textsc{J.~Shen},  \textsc{A.\,C. Kummel},  and
  \textsc{S.~Lee},
 \jr{Surf. Sci. Rep.} \textbf{66}, 1 (2011).


\bibitem{Abe:2007jm}
 \textsc{M.~Abe},  \textsc{Y.~Sugimoto},  \textsc{T.~Namikawa},
  \textsc{K.~Morita},  \textsc{N.~Oyabu},  and  \textsc{S.~Morita},
 \jr{Appl. Phys. Lett.} \textbf{90}, 203103 (2007).


\bibitem{Sugimoto:2008bs}
 \textsc{Y.~Sugimoto},  \textsc{T.~Namikawa},  \textsc{K.~Miki},
  \textsc{M.~Abe},  and  \textsc{S.~Morita},
 \jr{Phys. Rev. B} \textbf{77}, 195424 (2008).


\bibitem{Albers:2009ig}
 \textsc{B.\,J. Albers},  \textsc{T.\,C. Schwendemann},  \textsc{M.\,Z.
  Baykara},  \textsc{N.~Pilet},  \textsc{M.~Liebmann},  \textsc{E.\,I. Altman},
   and  \textsc{U.\,D. Schwarz},
 \jr{Nature Nanotechnology} \textbf{4}, 307 (2009).


\bibitem{Sugimoto:2012dc}
 \textsc{Y.~Sugimoto},  \textsc{K.~Ueda},  \textsc{M.~Abe},  and
  \textsc{S.~Morita},
 \jr{J. Phys.: Condens. Matter} \textbf{24}, 084008 (2012).


\bibitem{Grimme:2004}
 \textsc{S.~Grimme},
 \jr{J. Comput. Chem.} \textbf{25}, 1463 (2004).


\bibitem{Grimme:2006}
 \textsc{S.~Grimme},
 \jr{J. Comput. Chem.} \textbf{27}, 1787 (2006).


\bibitem{Grimme:2010}
 \textsc{S.~Grimme},  \textsc{J.~Antony},  \textsc{S.~Ehrlich},  and
  \textsc{H.~Krieg},
 \jr{J. Chem. Phys.} \textbf{132}, 154104 (2010).


\bibitem{Tkatchenko:2009}
 \textsc{A.~Tkatchenko} and  \textsc{M.~Scheffler},
 \jr{Phys. Rev. Lett.} \textbf{102}, 073005 (2009).


\bibitem{Dion:2004}
 \textsc{M.~Dion},  \textsc{H.~Rydberg},  \textsc{E.~Schroder},
  \textsc{D.~Langreth},  and  \textsc{B.~Lundqvist},
 \jr{Phys. Rev. Lett.} \textbf{92}, 246401 (2004).


\bibitem{Klimes:2009ei}
 \textsc{J.~Klime{\v s}},  \textsc{D.\,R. Bowler},  and
  \textsc{A.~Michaelides},
 \jr{J. Phys.: Condens. Matter} \textbf{22}, 022201 (2009).


\bibitem{Vanin:2010cr}
 \textsc{M.~Vanin},  \textsc{J.\,J. Mortensen},  \textsc{A.\,K. Kelkkanen},
  \textsc{J.\,M. Garcia-Lastra},  \textsc{K.\,S. Thygesen},  and
  \textsc{K.\,W. Jacobsen},
 \jr{Phys. Rev. B} \textbf{81}, 081408 (2010).


\bibitem{Tersoff:1985}
 \textsc{J.~Tersoff} and  \textsc{D.\,R. Hamann},
 \jr{Phys. Rev. B} \textbf{31}, 805 (1985).


\bibitem{Hofer:2003wk}
 \textsc{W.\,A. Hofer},  \textsc{A.\,S. Foster},  and  \textsc{A.\,L.
  Shluger},
 \jr{Rev. Mod. Phys.} \textbf{75}, 1287 (2003).


\bibitem{Kantorovich:2006ca}
 \textsc{L.~Kantorovich} and  \textsc{C.~Hobbs},
 \jr{Phys. Rev. B} \textbf{73}, 245420 (2006).


\bibitem{Chan:2009jb}
 \textsc{T.\,L. Chan},  \textsc{C.~Wang},  \textsc{K.~Ho},  and
  \textsc{J.~Chelikowsky},
 \jr{Phys. Rev. Lett.} \textbf{102}, 176101 (2009).


\bibitem{Klink:1995}
 \textsc{C.~Klink},  \textsc{I.~Stensgaard},  \textsc{F.~Besenbacher},  and
  \textsc{E.~Laegsgaard},
 \jr{Surf. Sci.} \textbf{342}, 250 (1995).


\bibitem{Dedkov:2008a}
 \textsc{Y.\,S. Dedkov},  \textsc{M.~Fonin},  \textsc{U.~R\"udiger},  and
  \textsc{C.~Laubschat},
 \jr{Phys. Rev. Lett.} \textbf{100}, 107602 (2008).


\bibitem{Eom:2009}
 \textsc{D.~Eom},  \textsc{D.~Prezzi},  \textsc{K.\,T. Rim},  \textsc{H.~Zhou},
   \textsc{M.~Lefenfeld},  \textsc{S.~Xiao},  \textsc{C.~Nuckolls},
  \textsc{M.\,S. Hybertsen},  \textsc{T.\,F. Heinz},  and  \textsc{G.\,W.
  Flynn},
 \jr{Nano Lett.} \textbf{9}, 2844 (2009).


\bibitem{Varykhalov:2009}
 \textsc{A.~Varykhalov} and  \textsc{O.~Rader},
 \jr{Phys. Rev. B} \textbf{80}, 035437 (2009).


\bibitem{Dedkov:2010jh}
 \textsc{Y.\,S. Dedkov} and  \textsc{M.~Fonin},
 \jr{New J. Phys.} \textbf{12}, 125004 (2010).


\bibitem{Lahiri:2011iu}
 \textsc{J.~Lahiri},  \textsc{T.~Miller},  \textsc{L.~Adamska},  \textsc{I.\,I.
  Oleynik},  and  \textsc{M.~Batzill},
 \jr{Nano Lett.} \textbf{11}, 518 (2011).


\bibitem{Dzemiantsova:2011bv}
 \textsc{L.\,V. Dzemiantsova},  \textsc{M.~Karolak},  \textsc{F.~Lofink},
  \textsc{A.~Kubetzka},  \textsc{B.~Sachs},  \textsc{K.~von Bergmann},
  \textsc{S.~Hankemeier},  \textsc{T.\,O. Wehling},  \textsc{R.~Fr{\"o}mter},
  \textsc{H.\,P. Oepen},  \textsc{A.\,I. Lichtenstein},  and
  \textsc{R.~Wiesendanger},
 \jr{Phys. Rev. B} \textbf{84}, 205431 (2011).


\bibitem{Jacobson:2012gv}
 \textsc{P.~Jacobson},  \textsc{B.~St{\"o}ger},  \textsc{A.~Garhofer},
  \textsc{G.\,S. Parkinson},  \textsc{M.~Schmid},  \textsc{R.~Caudillo},
  \textsc{F.~Mittendorfer},  \textsc{J.~Redinger},  and
  \textsc{U.~Diebold},
 \jr{J. Phys. Chem. Lett.} \textbf{3}, 136 (2012).


\bibitem{Jacobson:2012be}
 \textsc{P.~Jacobson},  \textsc{B.~St{\"o}ger},  \textsc{A.~Garhofer},
  \textsc{G.\,S. Parkinson},  \textsc{M.~Schmid},  \textsc{R.~Caudillo},
  \textsc{F.~Mittendorfer},  \textsc{J.~Redinger},  and
  \textsc{U.~Diebold},
 \jr{ACS Nano} \textbf{6}, 3564 (2012).


\bibitem{Bianchini:2014faa}
 \textsc{F.~Bianchini},  \textsc{L.\,L. Patera},  \textsc{M.~Peressi},
  \textsc{C.~Africh},  and  \textsc{G.~Comelli},
 \jr{J. Phys. Chem. Lett.} \textbf{5}, 467 (2014).


\bibitem{Prezzi:2014bk}
 \textsc{D.~Prezzi},  \textsc{D.~Eom},  \textsc{K.\,T. Rim},  \textsc{H.~Zhou},
   \textsc{S.~Xiao},  \textsc{C.~Nuckolls},  \textsc{T.\,F. Heinz},
  \textsc{G.\,W. Flynn},  and  \textsc{M.\,S. Hybertsen},
 \jr{ACS Nano} \textbf{8}, 5765 (2014).


\bibitem{Bertoni:2004}
 \textsc{G.~Bertoni},  \textsc{L.~Calmels},  \textsc{A.~Altibelli},  and
  \textsc{V.~Serin},
 \jr{Phys. Rev. B} \textbf{71}, 075402 (2004).


\bibitem{Weser:2011}
 \textsc{M.~Weser},  \textsc{E.\,N. Voloshina},  \textsc{K.~Horn},  and
  \textsc{Y.\,S. Dedkov},
 \jr{Phys. Chem. Chem. Phys.} \textbf{13}, 7534 (2011).


\bibitem{Voloshina:2011NJP}
 \textsc{E.\,N. Voloshina},  \textsc{A.~Generalov},  \textsc{M.~Weser},
  \textsc{S.~B{\"o}ttcher},  \textsc{K.~Horn},  and  \textsc{Y.\,S.
  Dedkov},
 \jr{New J. Phys.} \textbf{13}, 113028 (2011).


\bibitem{NDiaye:2008qq}
 \textsc{A.\,T. N'Diaye},  \textsc{J.~Coraux},  \textsc{T.\,N. Plasa},
  \textsc{C.~Busse},  and  \textsc{T.~Michely},
 \jr{New J. Phys.} \textbf{10}, 043033 (2008).


\bibitem{VazquezDeParga:2008}
 \textsc{A.\,L. V{\'a}zquez De~Parga},  \textsc{F.~Calleja},
  \textsc{B.~Borca},  \textsc{M.\,C.\,G. Passeggi},  \textsc{J.\,J. Hinarejos},
   \textsc{F.~Guinea},  and  \textsc{R.~Miranda},
 \jr{Phys. Rev. Lett.} \textbf{100}, 056807 (2008).


\bibitem{Stradi:2011be}
 \textsc{D.~Stradi},  \textsc{S.~Barja},  \textsc{C.~D{\'\i}az},
  \textsc{M.~Garnica},  \textsc{B.~Borca},  \textsc{J.~Hinarejos},
  \textsc{D.~S{\'a}nchez-Portal},  \textsc{M.~Alcam{\'\i}},  \textsc{A.~Arnau},
   \textsc{A.~V{\'a}zquez\,de Parga},  \textsc{R.~Miranda},  and
  \textsc{F.~Mart{\'\i}n},
 \jr{Phys. Rev. Lett.} \textbf{106}, 186102 (2011).


\bibitem{Stradi:2012hw}
 \textsc{D.~Stradi},  \textsc{S.~Barja},  \textsc{C.~D{\'\i}az},
  \textsc{M.~Garnica},  \textsc{B.~Borca},  \textsc{J.~Hinarejos},
  \textsc{D.~S{\'a}nchez-Portal},  \textsc{M.~Alcam{\'\i}},  \textsc{A.~Arnau},
   \textsc{A.\,L. V{\'a}zquez De~Parga},  \textsc{R.~Miranda},  and
  \textsc{F.~Mart{\'\i}n},
 \jr{Phys. Rev. B} \textbf{85}, 121404 (2012).


\bibitem{Voloshina:2012a}
 \textsc{E.\,N. Voloshina},  \textsc{Y.\,S. Dedkov},
  \textsc{S.~Torbr{\"u}gge},  \textsc{A.~Thissen},  and
  \textsc{M.~Fonin},
 \jr{Appl. Phys. Lett.} \textbf{100}, 241606 (2012).


\bibitem{Voloshina:2013dq}
 \textsc{E.\,N. Voloshina},  \textsc{E.~Fertitta},  \textsc{A.~Garhofer},
  \textsc{F.~Mittendorfer},  \textsc{M.~Fonin},  \textsc{A.~Thissen},  and
  \textsc{Y.\,S. Dedkov},
 \jr{Sci. Rep.} \textbf{3}, 1072 (2013).


\bibitem{Dedkov:2014di}
 \textsc{Y.~Dedkov} and  \textsc{E.~Voloshina},
 \jr{Phys. Chem. Chem. Phys.} \textbf{16}, 3894 (2014).


\bibitem{Busse:2011}
 \textsc{C.~Busse},  \textsc{P.~Lazic},  \textsc{R.~Djemour},
  \textsc{J.~Coraux},  \textsc{T.~Gerber},  \textsc{N.~Atodiresei},
  \textsc{V.~Caciuc},  \textsc{R.~Brako},  \textsc{A.\,T. N'Diaye},
  \textsc{S.~Bluegel},  \textsc{J.~Zegenhagen},  and
  \textsc{T.~Michely},
 \jr{Phys. Rev. Lett.} \textbf{107}, 036101 (2011).


\bibitem{Hamalainen:2013jj}
 \textsc{S.\,K. H{\"a}m{\"a}l{\"a}inen},  \textsc{M.\,P. Boneschanscher},
  \textsc{P.\,H. Jacobse},  \textsc{I.~Swart},  \textsc{K.~Pussi},
  \textsc{W.~Moritz},  \textsc{J.~Lahtinen},  \textsc{P.~Liljeroth},  and
  \textsc{J.~Sainio},
 \jr{Phys. Rev. B} \textbf{88}, 201406 (2013).


\bibitem{Vinogradov:2012fg}
 \textsc{N.~Vinogradov},  \textsc{A.~Zakharov},  \textsc{V.~Kocevski},
  \textsc{J.~Rusz},  \textsc{K.~Simonov},  \textsc{O.~Eriksson},
  \textsc{A.~Mikkelsen},  \textsc{E.~Lundgren},  \textsc{A.~Vinogradov},
  \textsc{N.~Martensson},  and  \textsc{A.~Preobrajenski},
 \jr{Phys. Rev. Lett.} \textbf{109}, 026101 (2012).


\bibitem{Boneschanscher:2012bg}
 \textsc{M.\,P.\,M. Boneschanscher},  \textsc{J.\,J. van\,der Lit},
  \textsc{Z.\,Z. Sun},  \textsc{I.\,I. Swart},  \textsc{P.\,P. Liljeroth},  and
   \textsc{D.\,D. Vanmaekelbergh},
 \jr{ACS Nano} \textbf{6}, 10216 (2012).


\bibitem{Koch:2013bk}
 \textsc{S.~Koch},  \textsc{D.~Stradi},  \textsc{E.~Gnecco},
  \textsc{S.~Barja},  \textsc{S.~Kawai},  \textsc{C.~D{\'\i}az},
  \textsc{M.~Alcam{\'\i}},  \textsc{F.~Mart{\'\i}n},  \textsc{A.\,L.\,V.
  de~Parga},  \textsc{R.~Miranda},  \textsc{T.~Glatzel},  and
  \textsc{E.~Meyer},
 \jr{ACS Nano} \textbf{7}, 2927 (2013).


\bibitem{Wang:2010ky}
 \textsc{B.~Wang},  \textsc{M.~Caffio},  \textsc{C.~Bromley},
  \textsc{H.~Fr{\"u}chtl},  and  \textsc{R.~Schaub},
 \jr{ACS Nano} \textbf{4}, 5773 (2010).


\bibitem{Borca:2010jj}
 \textsc{B.~Borca},  \textsc{S.~Barja},  \textsc{M.~Garnica},
  \textsc{D.~S{\'a}nchez-Portal},  \textsc{V.~Silkin},  \textsc{E.~Chulkov},
  \textsc{C.~Hermanns},  \textsc{J.~Hinarejos},  \textsc{A.~V{\'a}zquez\,de
  Parga},  \textsc{A.~Arnau},  \textsc{P.~Echenique},  and
  \textsc{R.~Miranda},
 \jr{Phys. Rev. Lett.} \textbf{105}, 036804 (2010).


\bibitem{Wang:2010jw}
 \textsc{B.~Wang},  \textsc{S.~Gunther},  \textsc{J.~Wintterlin},  and
  \textsc{M.\,L. Bocquet},
 \jr{New J. Phys.} \textbf{12}, 043041 (2010).


\bibitem{Altenburg:2014cq}
 \textsc{S.\,J. Altenburg} and  \textsc{R.~Berndt},
 \jr{New J. Phys.} \textbf{16}, 053036 (2014).


\bibitem{Zhang:2011ky}
 \textsc{H.~Zhang},  \textsc{J.~Sun},  \textsc{T.~Low},  \textsc{L.~Zhang},
  \textsc{Y.~Pan},  \textsc{Q.~Liu},  \textsc{J.~Mao},  \textsc{H.~Zhou},
  \textsc{H.~Guo},  \textsc{S.~Du},  \textsc{F.~Guinea},  and  \textsc{H.\,J.
  Gao},
 \jr{Phys. Rev. B} \textbf{84}, 245436 (2011).


\bibitem{Zhang:2012il}
 \textsc{H.~Zhang},  \textsc{W.\,D. Xiao},  \textsc{J.~Mao},  \textsc{H.~Zhou},
   \textsc{G.~Li},  \textsc{Y.~Zhang},  \textsc{L.~Liu},  \textsc{S.~Du},  and
  \textsc{H.\,J. Gao},
 \jr{J. Phys. Chem. C} \textbf{116}, 11091 (2012).


\bibitem{Yang:2012kq}
 \textsc{K.~Yang},  \textsc{W.\,D. Xiao},  \textsc{Y.\,H. Jiang},
  \textsc{H.\,G. Zhang},  \textsc{L.\,W. Liu},  \textsc{J.\,H. Mao},
  \textsc{H.\,T. Zhou},  \textsc{S.\,X. Du},  and  \textsc{H.\,J. Gao},
 \jr{J. Phys. Chem. C} \textbf{116}, 14052 (2012).


\bibitem{Li:2012in}
 \textsc{G.~Li},  \textsc{H.\,T. Zhou},  \textsc{L.\,D. Pan},
  \textsc{Y.~Zhang},  \textsc{J.\,H. Mao},  \textsc{Q.~Zou},  \textsc{H.\,M.
  Guo},  \textsc{Y.\,L. Wang},  \textsc{S.\,X. Du},  and  \textsc{H.\,J.
  Gao},
 \jr{Appl. Phys. Lett.} \textbf{100}, 013304 (2012).


\bibitem{Hamalainen:2012bx}
 \textsc{S.\,K. H{\"a}m{\"a}l{\"a}inen},  \textsc{M.~Stepanova},
  \textsc{R.~Drost},  \textsc{P.~Liljeroth},  \textsc{J.~Lahtinen},  and
  \textsc{J.~Sainio},
 \jr{J. Phys. Chem. C} \textbf{116}, 20433 (2012).


\bibitem{Jarvinen:2014va}
 \textsc{P.~J{\"a}rvinen},  \textsc{S.\,K. H{\"a}m{\"a}l{\"a}inen},
  \textsc{M.~Ij{\"a}s},  \textsc{A.~Harju},  and  \textsc{P.~Liljeroth},
 \jr{J. Phys. Chem. C} \textbf{118}, 13320 (2014).


\bibitem{NDiaye:2009a}
 \textsc{A.\,T. N'Diaye},  \textsc{T.~Gerber},  \textsc{C.~Busse},
  \textsc{J.~Myslivecek},  \textsc{J.~Coraux},  and
  \textsc{T.~Michely},
 \jr{New J. Phys.} \textbf{11}, 103045 (2009).


\bibitem{Pan:2009}
 \textsc{Y.~Pan},  \textsc{M.~Gao},  \textsc{L.~Huang},  \textsc{F.~Liu},  and
  \textsc{H.\,J. Gao},
 \jr{Appl. Phys. Lett.} \textbf{95}, 093106 (2009).


\bibitem{Rusponi:2010}
 \textsc{S.~Rusponi},  \textsc{M.~Papagno},  \textsc{P.~Moras},
  \textsc{S.~Vlaic},  \textsc{M.~Etzkorn},  \textsc{P.~Sheverdyaeva},
  \textsc{D.~Pacil{\'e}},  \textsc{H.~Brune},  and  \textsc{C.~Carbone},
 \jr{Phys. Rev. Lett.} \textbf{105}, 246803 (2010).


\bibitem{VoVan:2011ia}
 \textsc{C.~Vo-Van},  \textsc{S.~Schumacher},  \textsc{J.~Coraux},
  \textsc{V.~Sessi},  \textsc{O.~Fruchart},  \textsc{N.\,B. Brookes},
  \textsc{P.~Ohresser},  and  \textsc{T.~Michely},
 \jr{Appl. Phys. Lett.} \textbf{99}, 142504 (2011).


\bibitem{Papagno:2012hl}
 \textsc{M.~Papagno},  \textsc{S.~Rusponi},  \textsc{P.\,M. Sheverdyaeva},
  \textsc{S.~Vlaic},  \textsc{M.~Etzkorn},  \textsc{D.~Pacil{\'e}},
  \textsc{P.~Moras},  \textsc{C.~Carbone},  and  \textsc{H.~Brune},
 \jr{ACS Nano} \textbf{6}, 199 (2012).


\bibitem{Knudsen:2012ei}
 \textsc{J.~Knudsen},  \textsc{P.~Feibelman},  \textsc{T.~Gerber},
  \textsc{E.~Gr{\aa}n{\"a}s},  \textsc{K.~Schulte},  \textsc{P.~Stratmann},
  \textsc{J.~Andersen},  and  \textsc{T.~Michely},
 \jr{Phys. Rev. B} \textbf{85}, 035407 (2012).


\bibitem{Ndiaye:2006}
 \textsc{A.~N'diaye},  \textsc{S.~Bleikamp},  \textsc{P.~Feibelman},  and
  \textsc{T.~Michely},
 \jr{Phys. Rev. Lett.} \textbf{97}, 215501 (2006).


\bibitem{Feibelman:2009}
 \textsc{P.~Feibelman},
 \jr{Phys. Rev. B} \textbf{80}, 085412 (2009).


\bibitem{Dresselhaus:2002}
 \textsc{M.~Dresselhaus} and  \textsc{G.~Dresselhaus},
 \jr{Adv. Phys.} \textbf{51}, 1 (2002).


\bibitem{Pacile:2013jc}
 \textsc{D.~Pacil{\'e}},  \textsc{P.~Leicht},  \textsc{M.~Papagno},
  \textsc{P.\,M. Sheverdyaeva},  \textsc{P.~Moras},  \textsc{C.~Carbone},
  \textsc{K.~Krausert},  \textsc{L.~Zielke},  \textsc{M.~Fonin},
  \textsc{Y.\,S. Dedkov},  \textsc{F.~Mittendorfer},  \textsc{J.~Doppler},
  \textsc{A.~Garhofer},  and  \textsc{J.~Redinger},
 \jr{Phys. Rev. B} \textbf{87}, 035420 (2013).


\bibitem{Vita:2014aa}
 \textsc{H.~Vita},  \textsc{S.~B{\"o}ttcher},  \textsc{K.~Horn},
  \textsc{E.\,N. Voloshina},  \textsc{R.\,E. Ovcharenko},  \textsc{T.~Kampen},
  \textsc{A.~Thissen},  and  \textsc{Y.\,S. Dedkov},
 \jr{Sci. Rep.} \textbf{4}, 5704 (2014).


\bibitem{Decker:2013ch}
 \textsc{R.~Decker},  \textsc{J.~Brede},  \textsc{N.~Atodiresei},
  \textsc{V.~Caciuc},  \textsc{S.~Bl{\"u}gel},  and
  \textsc{R.~Wiesendanger},
 \jr{Phys. Rev. B} \textbf{87}, 041403 (2013).


\bibitem{Bickel:2011fg}
 \textsc{J.\,E. Bickel},  \textsc{F.~Meier},  \textsc{J.~Brede},
  \textsc{A.~Kubetzka},  \textsc{K.~von Bergmann},  and
  \textsc{R.~Wiesendanger},
 \jr{Phys. Rev. B} \textbf{84}, 054454 (2011).


\bibitem{Varykhalov:2010a}
 \textsc{A.~Varykhalov},  \textsc{M.~Scholz},  \textsc{T.~Kim},  and
  \textsc{O.~Rader},
 \jr{Phys. Rev. B} \textbf{82}, 121101 (2010).


\bibitem{Walter:2011fj}
 \textsc{A.~Walter},  \textsc{S.~Nie},  \textsc{A.~Bostwick},  \textsc{K.\,S.
  Kim},  \textsc{L.~Moreschini},  \textsc{Y.\,J. Chang},
  \textsc{D.~Innocenti},  \textsc{K.~Horn},  \textsc{K.~McCarty},  and
  \textsc{E.~Rotenberg},
 \jr{Phys. Rev. B} \textbf{84}, 195443 (2011).


\bibitem{Schumacher:2013ge}
 \textsc{S.~Schumacher},  \textsc{D.\,F. F{\"o}rster},  \textsc{M.~R{\"o}sner},
   \textsc{T.\,O. Wehling},  and  \textsc{T.~Michely},
 \jr{Phys. Rev. Lett.} \textbf{110}, 086111 (2013).


\bibitem{Schumacher:2013hl}
 \textsc{S.~Schumacher},  \textsc{T.\,O. Wehling},  \textsc{P.~Lazic},
  \textsc{S.~Runte},  \textsc{D.\,F. F{\"o}rster},  \textsc{C.~Busse},
  \textsc{M.~Petrovi{\'c}},  \textsc{M.~Kralj},  \textsc{S.~Bl{\"u}gel},
  \textsc{N.~Atodiresei},  \textsc{V.~Caciuc},  and
  \textsc{T.~Michely},
 \jr{Nano Lett.} \textbf{13}, 5013 (2013).


\bibitem{Petrovic:2013vz}
 \textsc{M.~Petrovi{\'c}},  \textsc{I.\,S.\,r.\,R. cacute},  \textsc{S.~Runte},
   \textsc{C.~Busse},  \textsc{J.\,T. Sadowski},  \textsc{P.\,L. cacute},
  \textsc{I.\,P. cacute},  \textsc{Z.\,H. Pan},  \textsc{M.~Milun},
  \textsc{P.~Pervan},  \textsc{N.~Atodiresei},  \textsc{R.~Brako},
  \textsc{D.\,S.\,o.\,c.\,e. cacute},  \textsc{T.~Valla},  \textsc{T.~Michely},
   and  \textsc{M.~Kralj},
 \jr{Nature Communications} \textbf{4}, 2772 (2013).


\bibitem{Sicot:2012}
 \textsc{M.~Sicot},  \textsc{P.~Leicht},  \textsc{A.~Zusan},
  \textsc{S.~Bouvron},  \textsc{O.~Zander},  \textsc{M.~Weser},  \textsc{Y.\,S.
  Dedkov},  \textsc{K.~Horn},  and  \textsc{M.~Fonin},
 \jr{ACS Nano} \textbf{6}, 151 (2012).


\bibitem{Meng:2012ee}
 \textsc{L.~Meng},  \textsc{R.~Wu},  \textsc{H.~Zhou},  \textsc{G.~Li},
  \textsc{Y.~Zhang},  \textsc{L.~Li},  \textsc{Y.~Wang},  and  \textsc{H.\,J.
  Gao},
 \jr{Appl. Phys. Lett.} \textbf{100}, 083101 (2012).


\bibitem{Lizzit:2012hh}
 \textsc{S.~Lizzit},  \textsc{R.~Larciprete},  \textsc{P.~Lacovig},
  \textsc{M.~Dalmiglio},  \textsc{F.~Orlando},  \textsc{A.~Baraldi},
  \textsc{L.~Gammelgaard},  \textsc{L.~Barreto},  \textsc{M.~Bianchi},
  \textsc{E.~Perkins},  and  \textsc{P.~Hofmann},
 \jr{Nano Lett.} \textbf{12}, 4503 (2012).


\bibitem{Riedl:2010du}
 \textsc{C.~Riedl},  \textsc{C.~Coletti},  and  \textsc{U.~Starke},
 \jr{J. Phys. D: Appl. Phys.} \textbf{43}, 374009 (2010).


\bibitem{Feng:2012il}
 \textsc{X.~Feng},  \textsc{S.~Maier},  and  \textsc{M.~Salmeron},
 \jr{J. Am. Chem. Soc.} \textbf{134}, 5662 (2012).


\bibitem{Feng:2013bp}
 \textsc{X.~Feng},  \textsc{S.~Kwon},  \textsc{J.\,Y. Park},  and
  \textsc{M.~Salmeron},
 \jr{ACS Nano} \textbf{7}, 1718 (2013).


\bibitem{Granas:2012cf}
 \textsc{E.~Gr{\aa}n{\"a}s},  \textsc{J.~Knudsen},  \textsc{U.\,A.
  Schr{\"o}der},  \textsc{T.~Gerber},  \textsc{C.~Busse},  \textsc{M.\,A.
  Arman},  \textsc{K.~Schulte},  \textsc{J.\,N. Andersen},  and
  \textsc{T.~Michely},
 \jr{ACS Nano} \textbf{6}, 9951 (2012).


\bibitem{Larciprete:2012aaa}
 \textsc{R.~Larciprete},  \textsc{S.~Ulstrup},  \textsc{P.~Lacovig},
  \textsc{M.~Dalmiglio},  \textsc{M.~Bianchi},  \textsc{F.~Mazzola},
  \textsc{L.~Hornekaer},  \textsc{F.~Orlando},  \textsc{A.~Baraldi},
  \textsc{P.~Hofmann},  and  \textsc{S.~Lizzit},
 \jr{ACS Nano} \textbf{6}, 9551 (2012).


\bibitem{Jang:2013cn}
 \textsc{W.\,J. Jang},  \textsc{H.~Kim},  \textsc{J.\,H. Jeon},  \textsc{J.\,K.
  Yoon},  and  \textsc{S.\,J. Kahng},
 \jr{Phys. Chem. Chem. Phys.} \textbf{15}, 16019 (2013).


\bibitem{Granas:2013tl}
 \textsc{E.~Gr{\aa}n{\"a}s},  \textsc{M.~Andersen},  \textsc{M.\,A. Arman},
  \textsc{T.~Gerber},  \textsc{B.~Hammer},  \textsc{J.~Schnadt},
  \textsc{J.\,N. Andersen},  \textsc{T.~Michely},  and
  \textsc{J.~Knudsen},
 \jr{The Journal of Physical Chemistry C} \textbf{117}, 16438
  (2013).


\bibitem{Shikin:2000a}
 \textsc{A.~Shikin},  \textsc{Y.~Dedkov},  \textsc{V.~Adamchuk},
  \textsc{D.~Farias},  and  \textsc{K.~Rieder},
 \jr{Surf. Sci.} \textbf{452}, 1 (2000).


\bibitem{Varykhalov:2010}
 \textsc{A.~Varykhalov},  \textsc{W.~Gudat},  and  \textsc{O.~Rader},
 \jr{Adv. Mater.} \textbf{22}, 3307 (2010).


\bibitem{Cui:2012ki}
 \textsc{Y.~Cui},  \textsc{J.~Gao},  \textsc{L.~Jin},  \textsc{J.~Zhao},
  \textsc{D.~Tan},  \textsc{Q.~Fu},  and  \textsc{X.~Bao},
 \jr{Nano Res.} \textbf{5}, 352 (2012).


\bibitem{Zhang:2009qqq}
 \textsc{H.~Zhang},  \textsc{Q.~Fu},  \textsc{Y.~Cui},  \textsc{D.~Tan},  and
  \textsc{X.~Bao},
 \jr{J. Phys. Chem. C} \textbf{113}, 8296 (2009).


\bibitem{Zhang:2009a}
 \textsc{Y.~Zhang},  \textsc{T.\,T. Tang},  \textsc{C.~Girit},
  \textsc{Z.~Hao},  \textsc{M.\,C. Martin},  \textsc{A.~Zettl},  \textsc{M.\,F.
  Crommie},  \textsc{Y.\,R. Shen},  and  \textsc{F.~Wang},
 \jr{Nature} \textbf{459}, 820 (2009).


\bibitem{Sutter:2010a}
 \textsc{P.~Sutter},  \textsc{J.\,T. Sadowski},  and  \textsc{E.\,A.
  Sutter},
 \jr{J. Am. Chem. Soc.} \textbf{132}, 8175 (2010).


\bibitem{Kosynkin:2009bw}
 \textsc{D.\,V. Kosynkin},  \textsc{A.\,L. Higginbotham},
  \textsc{A.~Sinitskii},  \textsc{J.\,R. Lomeda},  \textsc{A.~Dimiev},
  \textsc{B.\,K. Price},  and  \textsc{J.\,M. Tour},
 \jr{Nature} \textbf{458}, 872 (2009).


\bibitem{Tao:2011kl}
 \textsc{C.~Tao},  \textsc{L.~Jiao},  \textsc{O.\,V. Yazyev},  \textsc{Y.\,C.
  Chen},  \textsc{J.~Feng},  \textsc{X.~Zhang},  \textsc{R.\,B. Capaz},
  \textsc{J.\,M. Tour},  \textsc{A.~Zettl},  \textsc{S.\,G. Louie},
  \textsc{H.~Dai},  and  \textsc{M.\,F. Crommie},
 \jr{Nature Physics} \textbf{7}, 616 (2011).


\bibitem{Nakada:1996us}
 \textsc{K.~Nakada},  \textsc{M.~Fujita},  \textsc{G.~Dresselhaus},  and
  \textsc{M.~Dresselhaus},
 \jr{Phys. Rev., B Condens. Matter} \textbf{54}, 17954 (1996).


\bibitem{Fujita:1996vs}
 \textsc{M.~Fujita},  \textsc{K.~Wakabayashi},  \textsc{K.~Nakada},  and
  \textsc{K.~Kusakabe},
 \jr{J. Phys. Soc. Jpn.} \textbf{65}, 1920 (1996).


\bibitem{Wakabayashi:1999ti}
 \textsc{K.~Wakabayashi},  \textsc{M.~Fujita},  \textsc{H.~Ajiki},  and
  \textsc{M.~Sigrist},
 \jr{Phys. Rev. B} \textbf{59}, 8271 (1999).


\bibitem{Ezawa:2006ic}
 \textsc{M.~Ezawa},
 \jr{Phys. Rev. B} \textbf{73}, 045432 (2006).


\bibitem{Brey:2006cb}
 \textsc{L.~Brey} and  \textsc{H.~Fertig},
 \jr{Phys. Rev. B} \textbf{73}, 235411 (2006).


\bibitem{Sasaki:2006wr}
 \textsc{K.Sasaki},  \textsc{S.~Murakami},  and  \textsc{R.~Saito},
 \jr{J. Phys. Soc. Jpn.} \textbf{75}, 074713 (2006).


\bibitem{Abanin:2006ds}
 \textsc{D.~Abanin},  \textsc{P.~Lee},  and  \textsc{L.~Levitov},
 \jr{Phys. Rev. Lett.} \textbf{96}, 176803 (2006).


\bibitem{Lee:2005ek}
 \textsc{H.~Lee},  \textsc{Y.\,W. Son},  \textsc{N.~Park},  \textsc{S.~Han},
  and  \textsc{J.~Yu},
 \jr{Phys. Rev. B} \textbf{72}, 174431 (2005).


\bibitem{Son:2006ky}
 \textsc{Y.\,W. Son},  \textsc{M.\,L. Cohen},  and  \textsc{S.\,G.
  Louie},
 \jr{Phys. Rev. Lett.} \textbf{97}, 216803 (2006).


\bibitem{Han:2007bl}
 \textsc{M.~Han},  \textsc{B.~Ozyilmaz},  \textsc{Y.~Zhang},  and
  \textsc{P.~Kim},
 \jr{Phys. Rev. Lett.} \textbf{98}, 206805 (2007).


\bibitem{Li:2008ht}
 \textsc{X.~Li},  \textsc{X.~Wang},  \textsc{L.~Zhang},  \textsc{S.~Lee},  and
  \textsc{H.~Dai},
 \jr{Science} \textbf{319}, 1229 (2008).


\bibitem{Wang:2010cf}
 \textsc{X.\,X. Wang} and  \textsc{H.\,H. Dai},
 \jr{Nat Chem} \textbf{2}, 661 (2010).


\bibitem{Datta:2008if}
 \textsc{S.\,S. Datta},  \textsc{D.\,R. Strachan},  \textsc{S.\,M. Khamis},
  and  \textsc{A.\,T.\,C. Johnson},
 \jr{Nano Lett.} \textbf{8}, 1912 (2008).


\bibitem{Ci:2008je}
 \textsc{L.~Ci},  \textsc{Z.~Xu},  \textsc{L.~Wang},  \textsc{W.~Gao},
  \textsc{F.~Ding},  \textsc{K.\,F. Kelly},  \textsc{B.\,I. Yakobson},  and
  \textsc{P.\,M. Ajayan},
 \jr{Nano Res.} \textbf{1}, 116 (2008).


\bibitem{Campos:2009cd}
 \textsc{L.\,C. Campos},  \textsc{V.\,R. Manfrinato},  \textsc{J.\,D.
  Sanchez-Yamagishi},  \textsc{J.~Kong},  and
  \textsc{P.~Jarillo-Herrero},
 \jr{Nano Lett.} \textbf{9}, 2600 (2009).


\bibitem{Pisani:2007ge}
 \textsc{L.~Pisani},  \textsc{J.~Chan},  \textsc{B.~Montanari},  and
  \textsc{N.~Harrison},
 \jr{Phys. Rev. B} \textbf{75}, 064418 (2007).


\bibitem{Saffarzadeh:2011fd}
 \textsc{A.~Saffarzadeh} and  \textsc{R.~Farghadan},
 \jr{Appl. Phys. Lett.} \textbf{98}, 023106 (2011).


\bibitem{Fu:2012de}
 \textsc{H.\,H. Fu} and  \textsc{K.\,l. Yao},
 \jr{Appl. Phys. Lett.} \textbf{100}, 013502 (2012).


\bibitem{Farghadan:2012gj}
 \textsc{R.~Farghadan} and  \textsc{E.~Saievar-Iranizad},
 \jr{J. Appl. Phys.} \textbf{111}, 014304 (2012).


\bibitem{Hawkins:2012kv}
 \textsc{P.~Hawkins},  \textsc{M.~Begliarbekov},  \textsc{M.~Zivkovic},
  \textsc{S.~Strauf},  and  \textsc{C.\,P. Search},
 \jr{J. Phys. Chem. C} \textbf{116}, 18382 (2012).


\bibitem{Chen:2013fa}
 \textsc{Y.\,C. Chen},  \textsc{D.\,G. de~Oteyza},  \textsc{Z.~Pedramrazi},
  \textsc{C.~Chen},  \textsc{F.\,R. Fischer},  and  \textsc{M.\,F.
  Crommie},
 \jr{ACS Nano} \textbf{7}, 6123 (2013).


\bibitem{Liang:2012wi}
 \textsc{L.~Liang} and  \textsc{V.~Meunier},
 \jr{Phys. Rev. B} \textbf{86}, 195404 (2012).


\bibitem{Jiao:2010de}
 \textsc{L.~Jiao},  \textsc{X.~Wang},  \textsc{G.~Diankov},  \textsc{H.~Wang},
  and  \textsc{H.~Dai},
 \jr{Nature Nanotechnology} \textbf{5}, 321 (2010).


\bibitem{Feldner:2011cm}
 \textsc{H.~Feldner},  \textsc{Z.\,Y. Meng},  \textsc{T.\,C. Lang},
  \textsc{F.\,F. Assaad},  \textsc{S.~Wessel},  and
  \textsc{A.~Honecker},
 \jr{Phys. Rev. Lett.} \textbf{106}, 226401 (2011).


\bibitem{Steinkasserer:2014iv}
 \textsc{L.\,E.\,M. Steinkasserer},  \textsc{B.~Paulus},  and
  \textsc{E.~Voloshina},
 \jr{Chem. Phys. Lett.} \textbf{597}, 148 (2014).


\bibitem{Ritter:2009hw}
 \textsc{K.\,A. Ritter} and  \textsc{J.\,W. Lyding},
 \jr{Nature Materials} \textbf{8}, 235 (2009).


\bibitem{Pan:2012hp}
 \textsc{M.~Pan},  \textsc{E.\,C. Gir{\~a}o},  \textsc{X.~Jia},
  \textsc{S.~Bhaviripudi},  \textsc{Q.~Li},  \textsc{J.~Kong},
  \textsc{V.~Meunier},  and  \textsc{M.\,S. Dresselhaus},
 \jr{Nano Lett.} \textbf{12}, 1928 (2012).


\bibitem{Li:2013ih}
 \textsc{Y.~Li},  \textsc{W.~Zhang},  \textsc{M.~Morgenstern},  and
  \textsc{R.~Mazzarello},
 \jr{Phys. Rev. Lett.} \textbf{110}, 216804 (2013).


\bibitem{Archambault:2013fd}
 \textsc{C.~Archambault} and  \textsc{A.~Rochefort},
 \jr{ACS Nano} \textbf{7}, 5414 (2013).


\bibitem{Trauzettel:2007}
 \textsc{B.~Trauzettel},  \textsc{D.\,V. Bulaev},  \textsc{D.~Loss},  and
  \textsc{G.~Burkard},
 \jr{Nature Physics} \textbf{3}, 192 (2007).


\bibitem{DasSarma:2011br}
 \textsc{S.~Das~Sarma},  \textsc{S.~Adam},  \textsc{E.\,H. Hwang},  and
  \textsc{E.~Rossi},
 \jr{Rev. Mod. Phys.} \textbf{83}, 407 (2011).


\bibitem{Rycerz:2007}
 \textsc{A.~Rycerz},  \textsc{J.~Tworzydlo},  and  \textsc{C.\,W.\,J.
  Beenakker},
 \jr{Nature Physics} \textbf{3}, 172 (2007).


\bibitem{Ponomarenko:2008ej}
 \textsc{L.\,A. Ponomarenko},  \textsc{F.~Schedin},  \textsc{M.\,I.
  Katsnelson},  \textsc{R.~Yang},  \textsc{E.\,W. Hill},  \textsc{K.\,S.
  Novoselov},  and  \textsc{A.\,K. Geim},
 \jr{Science} \textbf{320}, 356 (2008).


\bibitem{Hamalainen:2011ja}
 \textsc{S.\,K. H{\"a}m{\"a}l{\"a}inen},  \textsc{Z.~Sun},  \textsc{M.\,P.
  Boneschanscher},  \textsc{A.~Uppstu},  \textsc{M.~Ij{\"a}s},
  \textsc{A.~Harju},  \textsc{D.~Vanmaekelbergh},  and
  \textsc{P.~Liljeroth},
 \jr{Phys. Rev. Lett.} \textbf{107}, 236803 (2011).


\bibitem{Coraux:2009}
 \textsc{J.~Coraux},  \textsc{A.\,T. N'Diaye},  \textsc{M.~Engler},
  \textsc{C.~Busse},  \textsc{D.~Wall},  \textsc{N.~Buckanie},
  \textsc{F.\,J.\,M.\,z. Heringdorf},  \textsc{R.\,v. Gastel},
  \textsc{B.~Poelsema},  and  \textsc{T.~Michely},
 \jr{New J. Phys.} \textbf{11}, 023006 (2009).


\bibitem{Lu:2011bg}
 \textsc{J.~Lu},  \textsc{P.\,S.\,E. Yeo},  \textsc{C.\,K. Gan},
  \textsc{P.~Wu},  and  \textsc{K.\,P. Loh},
 \jr{Nature Nanotech.} \textbf{6}, 247 (2011).


\bibitem{Forster:2012aa}
 \textsc{D.\,F. F{\"o}rster},  \textsc{T.\,O. Wehling},
  \textsc{S.~Schumacher},  \textsc{A.~Rosch},  and  \textsc{T.~Michely},
 \jr{New J. Phys.} \textbf{14}, 023022 (2012).


\bibitem{Starodub:2012cb}
 \textsc{E.~Starodub},  \textsc{N.\,C. Bartelt},  and  \textsc{K.\,F.
  McCarty},
 \jr{Appl. Phys. Lett.} \textbf{100}, 181604 (2012).


\bibitem{Jacobs:1998vo}
 \textsc{H.\,O. Jacobs},  \textsc{P.~Leuchtmann},  \textsc{O.\,J. Homan},  and
  \textsc{A.~Stemmer},
 \jr{J. Appl. Phys.} \textbf{84}, 1168 (1998).


\bibitem{Baier:2012ey}
 \textsc{R.~Baier},  \textsc{C.~Leendertz},  \textsc{M.~Lux-Steiner},  and
  \textsc{S.~Sadewasser},
 \jr{Phys. Rev. B} \textbf{85}, 165436 (2012).


\bibitem{Phark:2011de}
 \textsc{S.\,h. Phark},  \textsc{J.~Borme},  \textsc{A.\,L. Vanegas},
  \textsc{M.~Corbetta},  \textsc{D.~Sander},  and
  \textsc{J.~Kirschner},
 \jr{ACS Nano} \textbf{5}, 8162 (2011).


\bibitem{Lacovig:2009}
 \textsc{P.~Lacovig},  \textsc{M.~Pozzo},  \textsc{D.~Alf{\`e}},
  \textsc{P.~Vilmercati},  \textsc{A.~Baraldi},  and
  \textsc{S.~Lizzit},
 \jr{Phys. Rev. Lett.} \textbf{103}, 166101 (2009).


\bibitem{Enderlein:2010}
 \textsc{C.~Enderlein},  \textsc{Y.\,S. Kim},  \textsc{A.~Bostwick},
  \textsc{E.~Rotenberg},  and  \textsc{K.~Horn},
 \jr{New J. Phys.} \textbf{12}, 033014 (2010).


\bibitem{Jolie:2014ev}
 \textsc{W.~Jolie},  \textsc{F.~Craes},  \textsc{M.~Petrovi{\'c}},
  \textsc{N.~Atodiresei},  \textsc{V.~Caciuc},  \textsc{S.~Bl{\"u}gel},
  \textsc{M.~Kralj},  \textsc{T.~Michely},  and  \textsc{C.~Busse},
 \jr{Phys. Rev. B} \textbf{89}, 155435 (2014).


\bibitem{GarciaLekue:2014ce}
 \textsc{A.~Garcia-Lekue},  \textsc{T.~Balashov},  \textsc{M.~Olle},
  \textsc{G.~Ceballos},  \textsc{A.~Arnau},  \textsc{P.~Gambardella},
  \textsc{D.~S{\'a}nchez-Portal},  and  \textsc{A.~Mugarza},
 \jr{Phys. Rev. Lett.} \textbf{112}, 066802 (2014).


\bibitem{Levy:2010}
 \textsc{N.~Levy},  \textsc{S.\,A. Burke},  \textsc{K.\,L. Meaker},
  \textsc{M.~Panlasigui},  \textsc{A.~Zettl},  \textsc{F.~Guinea},
  \textsc{A.\,H.\,C. Neto},  and  \textsc{M.\,F. Crommie},
 \jr{Science} \textbf{329}, 544 (2010).


\bibitem{Lu:2014dn}
 \textsc{J.~Lu},  \textsc{A.\,H.\,C. Neto},  and  \textsc{K.\,P. Loh},
 \jr{Nature Communications} \textbf{3}, 823 (2014).


\end{thebibliography}

\providecommand{\WileyBibTextsc}{}
\let\textsc\WileyBibTextsc
\providecommand{\othercit}{}
\providecommand{\jr}[1]{#1}
\providecommand{\etal}{~et~al.}

\end{document}